\documentclass[arma,reqno]{svjour}
\makeatletter
\def\labelitemii{$\m@th\cdot$}
\makeatother
\long\def\drop#1{}

\let\limits\relax
\def\modd#1{\left|#1\right|}
\def\Modd#1{\left\|#1\right\|}

\def\Dual#1#2#3#4{{}_{#1}\langle #2,#3\rangle_{#4}}

\usepackage{amsmath,amssymb}

\usepackage{graphicx}
\usepackage{geompsfi, pstricks}

\usepackage{bm} % bold math
\def\AUTO{\texttt{AUTO}}

\newcommand{\pijl}{\longrightarrow}

\newcommand{\R}{{\mathbb R}}
\newcommand{\N}{{\mathbb N}}
\newcommand{\supp}{\mathrm{supp\ }}
\DeclareMathOperator\Int{Int}
\newcommand{\modu}[1]{\ (\text{mod } #1)}
\newcommand{\ep}{\varepsilon}
\newcommand{\ta}{\theta}

\newcommand{\V}{V}

\newcommand{\z}{\zeta}
\newcommand{\g}{\bar g}
\def\r{\bm{r}}
\def\d#1{\bm{d}_{#1}}
\def\e#1{\bm e_{#1}}
\def\pref#1{(\ref{#1})}

\def\arraystretch{1.1}

\title{Self-contact for rods on cylinders}

\authorrunning{Van der Heijden, Peletier, and Planqu\'e}

\author{G.~H.~M.~van der Heijden \and M.~A.~Peletier \and R.~Planqu\'e}
\begin{document}
\maketitle

\begin{abstract}
We study self-contact phenomena in elastic rods that are constrained
to lie on a cylinder. By choosing a particular set of variables to
describe the rod centerline the variational setting is made
particularly simple: the strain energy is a second-order functional of
a single scalar variable, and the self-contact constraint is written
as an integral inequality.

Using techniques from ode theory (comparison principles) and
variational calculus (cut-and-paste arguments) we fully characterize
the structure of constrained minimizers. An important auxiliary result
states that the set of self-contact points is continuous, a result
that contrasts with known examples from contact problems in free rods.
\keywords{elastic rods, calculus of variations, constrained minimization, 
self-contact, comparison principle.}\\ {\bf 2000 Mathematics Subject
Classification:} 34C11, 34C25, 34C60, 47J30, 49J40, 74G25, 74G55,
74G65.
\end{abstract}

\section{Introduction} 

The study of self-contact in elastic rods has seen some remarkable
progress over the last ten years, with highlights such as the
numerical work of Tobias, Coleman, and
Swigon~\cite{tobias.00,coleman.00,coleman.00a}, the introduction of
global curvature by Gonzalez and co-workers~\cite{gonzalez.01}, and
the derivation of the Euler-Lagrange equations for energy minimization
by Schuricht and Von der Mosel~\cite{schuricht.03}. Parallel advances
have been made on the highly related ideal knots and Gehring links,
where ropelength is minimized instead of elastic
energy~\cite{cantarella.02a,schuricht.04,cantarella.preprint.04}.

Despite this progress important questions remain open.  
We are still far from understanding
analytically the solutions of the Euler-Lagrange equations
for general contact situations. Even if we limit
ourselves to global minimizers of an appropriate energy functional, we
can prove little about the form of solutions as soon as contact is
taken into account.

For instance, a long-standing conjecture for closed elastic rods is
that in the limit of long rods under constant twist the global energy
minimizer should be a ply (double helix) with a loop on each end. 
If a structure of this type is
assumed, then the limiting pitch angle can be
determined~\cite{thompson.02}; but the difficult part actually
consists in showing that global minimizers have this
structure. Incidentally, since local minimizers of different type
have been found numerically~\cite{coleman.00,coleman.00a},
the restriction to global minimizers appears to be essential.

This example is typical for the current state of understanding: if
assumptions are made on the set of contacts, then characterizations
are possible~\cite{fraser.98,stump.01,thompson.02,heijden.03a}, but
for unrestrained geometry little is known rigorously. It shows how our
lack of understanding of energy minimizers is intimately linked to the
lack of knowledge about structure of the contact set. Examples show
that this structure can be non-trivial: for instance, non-contiguous
contact appears at the end of a ply in an elastic
rod~\cite{coleman.00a}.

\medskip

In this paper we study a problem of self-contact of an elastic rod in
which the rod has reduced freedom of movement: the centerline of the
rod is constrained to lie on the surface of a cylinder
(Figure~\ref{fig:cylcoord}). In contrast to the full three-dimensional
problem referred to above, the reduced dimensionality of this problem
enables us to give a near-complete characterization of global
minimizers, without making any \emph{a priori} assumptions on the
structure of the contact set. Notwithstanding this, determining the
structure of the contact set is a central element of this paper.

\begin{figure}[ht]
\pspicture(0,0)(11,3.7)
%\psgrid(0,0)(11,3.7)
\rput(5.7,2){
\includegraphics[height=2.9cm]{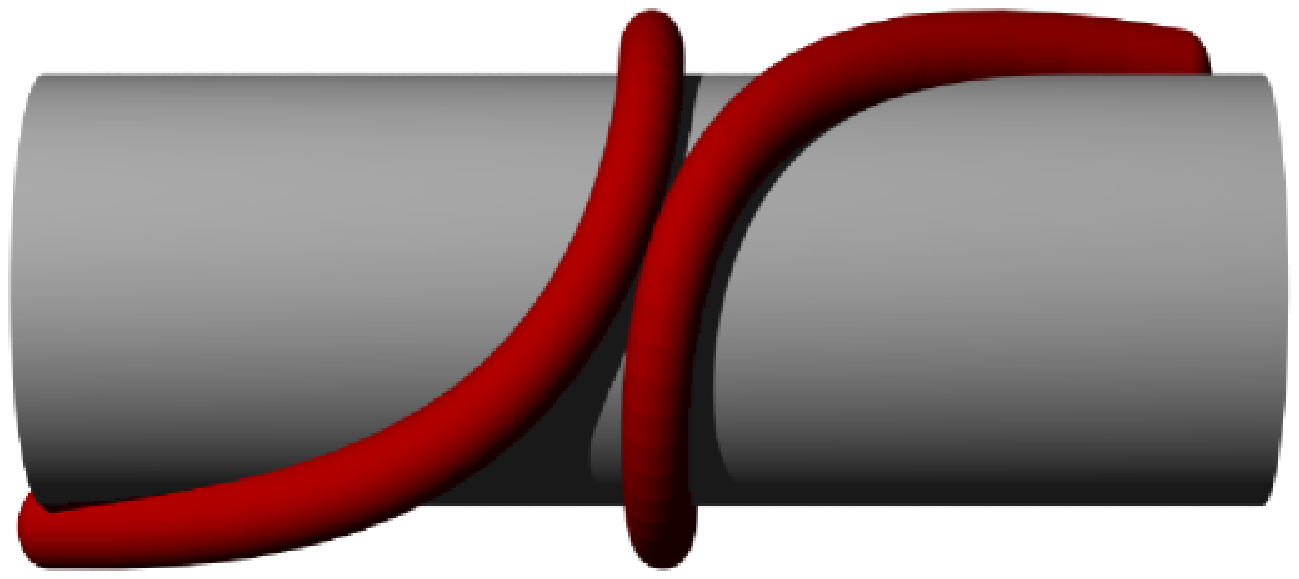}
}
\psline{->}(4,.5)(7,.5)
\uput[d](5.5,.5){$\zeta(x)$}
\pscurve{->}(2.5,1)(2.4,2)(2.5,3)
\uput[l](2.4,2){$x$}

%\psline{->}(6.3,3.5)(5.9,2.7)
%\uput[r](6.3,3.5){contact point}
\psline{->}(4.6,3.3)(5.65,2.45)
\uput[l](5.2,3.5){contact point}

\endpspicture
\caption{The centerline of a rod on a cylinder is described using
cylindrical coordinates: the independent variable $x$ is the
tangential coordinate, and the position of the centerline is given by
the function $\zeta(x)$ measuring distance along the cylinder axis.}
\label{fig:cylcoord}
\end{figure}

\drop{
\begin{figure}[ht]
\pspicture(0,0)(11,4)
%\psgrid(0,0)(11,4)
\rput(5.7,2){
\includegraphics[height=2.9cm]{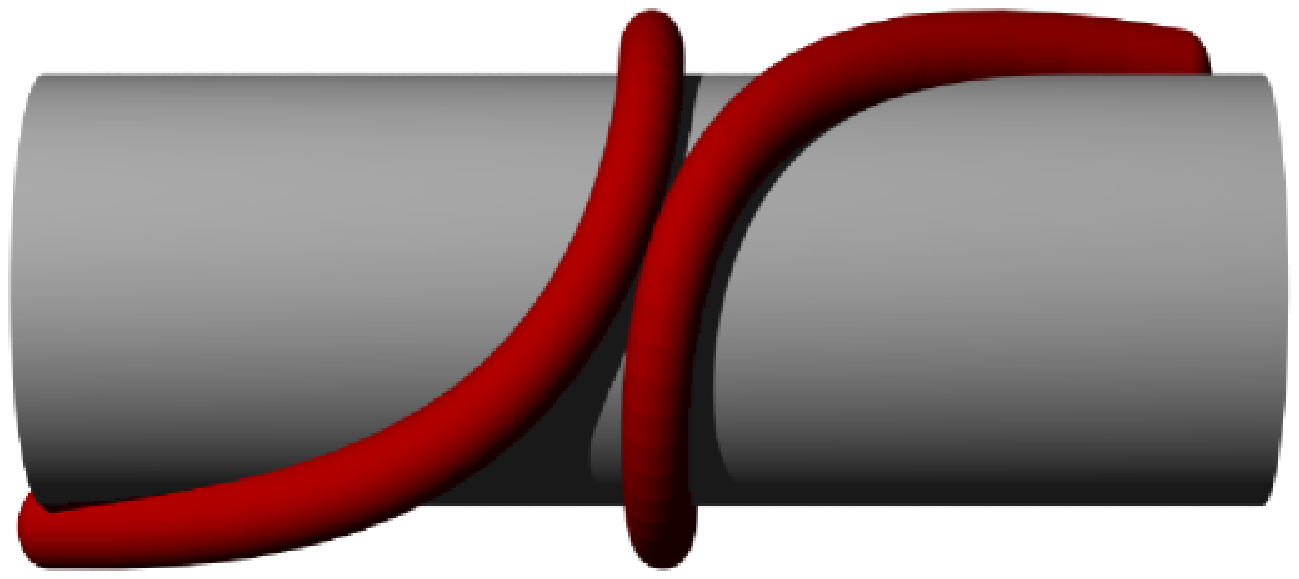}
}
%\psline{->}(4,.5)(7,.5)
%\uput[d](5.5,.5){$z(x)$}
%\pscurve{->}(2.5,1)(2.4,2)(2.5,3)
%\uput[l](2.4,2){$x$}

%\psline{->}(6.3,3.5)(5.9,2.7)
%\uput[l](6.3,3.5){contact point}

\endpspicture
\caption{
%The centerline of a rod on a cylinder is described using
%cylindrical coordinates $(z,x)$.
The rod is constrained to lie on the surface of a cylinder;
self-penetration is not allowed.
}
\label{fig:cylcoord}
\end{figure}
}

We transform the classical Cosserat model of an elastic, unshearable
rod of circular cross-section into a more convenient form. The
functional that is to be minimized (representing stored energy and
work done by the end moment) is
\[
F(u) = \int_0^T [\,a(u){u'}^2 + b(u)\,]\, ,
\]
where 
\begin{equation}
\label{def:ab}
a(u) =\frac 1{4\pi^2}\, \frac1 {(1+u^2)^{5/2}} \quad\text{and}\quad
b(u) = \frac 1{r^2(1+u^2)^{3/2}} - \frac{2M}{Br}\,\frac{\sqrt{1+u^2} -
u}{\sqrt{1+u^2}}.
\end{equation}
Here $r$ is the radius of the cylinder, $M$ the moment applied to the
end of the rod, and $B$ is the bending coefficient of the rod.  The
centerline of the rod is characterized by $\zeta(x)$, which measures
distance along the cylinder axis as a function of a tangential
independent variable $x$.  The unknown in this minimization problem is
the derivative $u(x) = \zeta'(x)$, which may by thought of as the
cotangent of the angle between the centerline tangent and the cylinder
axis; $u$ is zero when the rod curls around the cylinder orthogonal to
the axis, and $u=\pm\infty$ when the rod is parallel to the axis. This
transformation is detailed in Section~\ref{sec:derivation}.

The most interesting part of the variational problem is the
transformed contact condition (condition of non-self-penetration).  In
this paper we take the thickness of the rod to be zero; then the
non-self-penetration condition is
\begin{equation}
\label{cond:constraint}
\int_x^{x+1} u \geq 0 \qquad \text{for all}\quad 0\leq x\leq T-1,
\end{equation}
where the interval $[x,x+1]$ corresponds to one full turn around the
cylinder; this condition formalizes the intuitive idea that
non-self-penetration is equivalent to the condition `that the rod
remain on the same side of itself'.  This condition on $u$ makes the
variational problem a non-local obstacle problem.  Non-zero thickness
requires a contact condition that is substantially more involved
than~\pref{cond:constraint}; we comment on this situation in
Section~\ref{subsec:thickness}.

\medskip
%Apart from questions of existence and regularity of solutions
%of this minimization problem, which we address first,
%As a prelude to the study of the structure of solutions
%we first address 
%the issues of existence of minimizers of $F$ under the constraint~\pref{cond:constraint}, 
%and we derive
%the conditions of stationarity satisfied by these minimizers.
Both the background in rod theory and the independent mathematical
context of this minimization problem raise questions about the
solutions:
\begin{enumerate}
\item Do solutions exist?
\item What is the minimal, and what is the maximal regularity of minimizers?
\item When is there contact, i.e., when is the contact set
\begin{equation}
\label{def:omega_c}
\omega _c := \left\{ \,x \in [0,T-1]\ : \ \int_x^{x+1} u = 0\,\right\}
\end{equation}
non-empty?
\item Given that $\omega_c\not=\emptyset$, what is the structure of
$\omega_c$? Is the contact simply contained in a single interval,
or is the structure more intricate, as in the
examples of contact--skip--contact 
at the end of a ply~\cite{coleman.00} % Coleman & Swigon 2000
and in a (ropelength minimizing) 
clasp~\cite{cantarella.preprint.04}? % arXiv: math.DG/0402212

\item What form do the contact forces take?

\item Does the solution inherit the symmetry of the formulation? This is the
case for a symmetric rod on a cylinder without contact
condition~\cite{heijden.01.3}, but need not be true when taking
contact into account.

\end{enumerate}
In the rest of this paper we address these questions.

\section{Results}

The first main result of this investigation
(Theorem~\ref{thm:minimizing_rods_intersect}) shows that the contact
condition~\pref{cond:constraint} is essential: without this condition
the centerline of a rod will intersect itself.  A little experiment
with some string wrapped around a pencil will convince the reader that
this is the case. We also prove the regularity result that a
constrained minimizer~$u$ is of class $W^{2,\infty}$, and we derive
the Euler-Lagrange equation
\begin{equation}
\label{eq:EL-intro}
N(u)(x) := -2a(u(x))u''(x) -a'(u(x)){u'}^2(x) +b'(u(x)) = \int_{x-1}^x f,
\end{equation}
where the Lagrange multiplier $f$ is a non-negative Radon measure with
support contained in the contact set $\omega_c$
(Theorem~\ref{thm:EL-equations}).

\medskip

\begin{figure}[ht]
\pspicture(11,9)
\rput(5.5,7){
\centerline{\psfig{figure=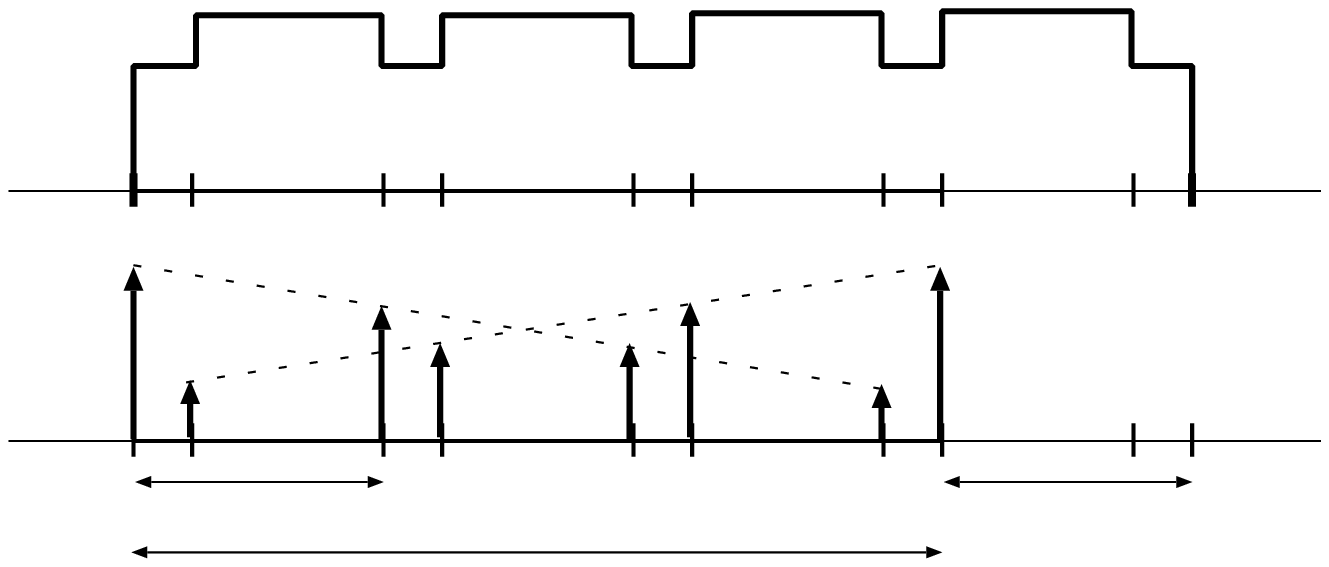,width=0.9\textwidth}}
}
\rput(5.5,2.5){
\centerline{\psfig{figure=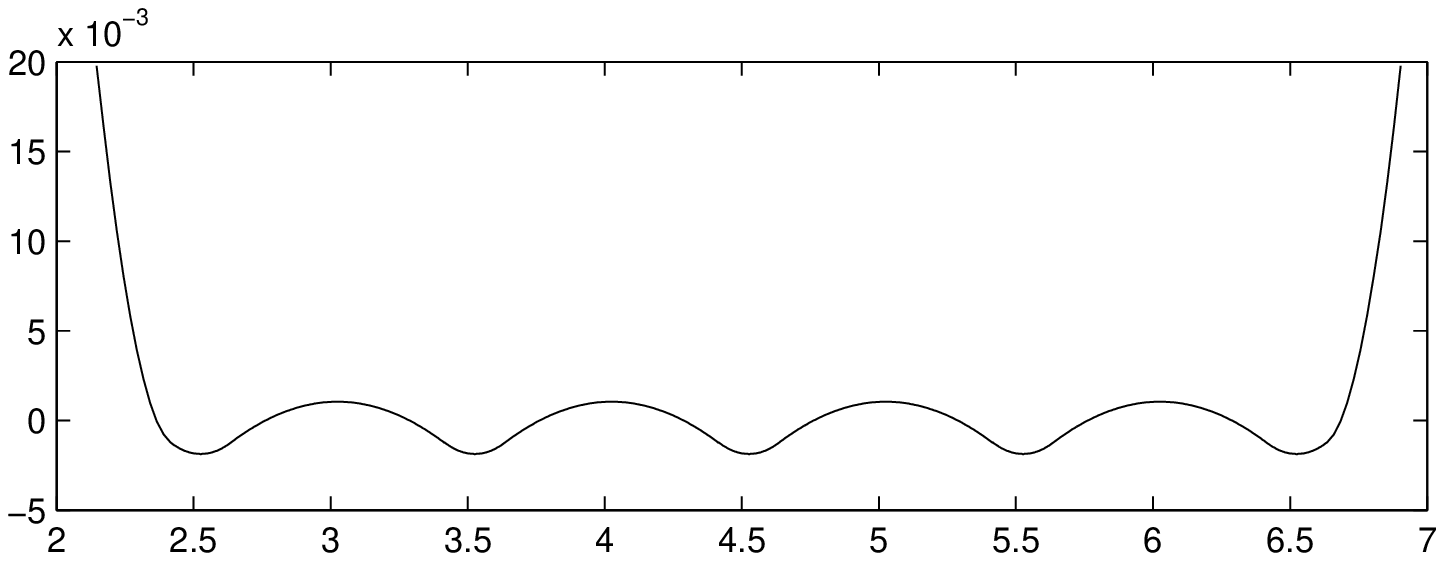,width=0.9\textwidth}}
}
\endpspicture
\caption{The function $g(x) = \int_{x-1}^x f$ is piecewise constant (top); 
the jumps correspond to Dirac delta functions in $f$ (middle). Note
that the support of $g$ is the set $\omega_c + [0,1]$ by the
definition of $g$. The solution $u$ corresponding to $f$ and $g$ is
shown at the bottom.}
\label{fig:fg}
\end{figure}

From stationarity alone, which is the basis of
Theorem~\ref{thm:EL-equations}, the characterization of $f$ as a
positive Radon measure appears to be optimal; no further information
can be extracted.  In Section~\ref{sec:interval} we use two different
additional assumptions to further characterize the contact set and
subsequently the measure $f$. In both cases we obtain the important
result that the contact set is a (possibly empty) interval and that
the measure $f$ is a sum of Dirac delta functions, as represented
schematically in Figure~\ref{fig:fg}.  The weighting of the delta
functions is shown in the middle of Figure~\ref{fig:fg}: there is a
linear decrease or increase in weight from one side of the contact set
to the other (Theorem~\ref{thm:structure_g_3}).  Since $f$ may be
interpreted as the contact force, we deduce that
\begin{itemize}
\item The contact force is concentrated in at most two tangential positions
$x_1$ and $x_2$, and in integer translates of $x_{1,2}$; 
\item The magnitude of the contact force is maximal at the contact
point where the rod lifts off, and decreases linearly with each turn.
Figure~\ref{fig:typical} graphically illustrates this behaviour.
\end{itemize}
The decrease in contact force with each turn can be understood in the
following way.  The difference between the contact forces on either
side of the rod creates a resulting force exerted on the rod, and the
two resultant forces that act at $x_{1,2}\mathbin{\mathrm{mod}} 1$
point in opposite directions.  If we imagine a single, closed ring
with two forces acting on it in this way, the two forces create a
moment that will bend the ring. This also happens with the coil of the
current problem, as is demonstrated by the small but definite
oscillations in the numerical solutions calculated in
Section~\ref{sec:numerics}.

\def\figsize{4cm}
\begin{figure}[!thbp]
\pspicture(11,5)
%linker twee plaatjes
\rput(3,2.5) {
\centerline{$\vcenter{\halign{#\cr
\includegraphics[width=\figsize]{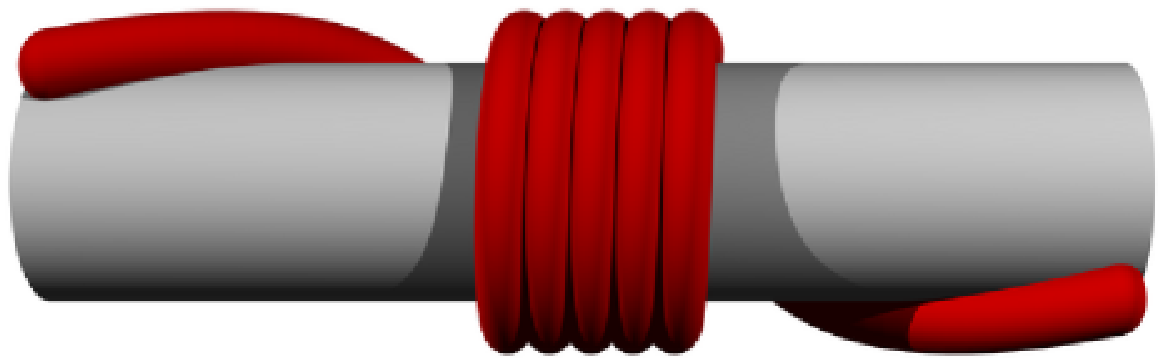}\cr\noalign{\vskip5mm}
\includegraphics[width=\figsize]{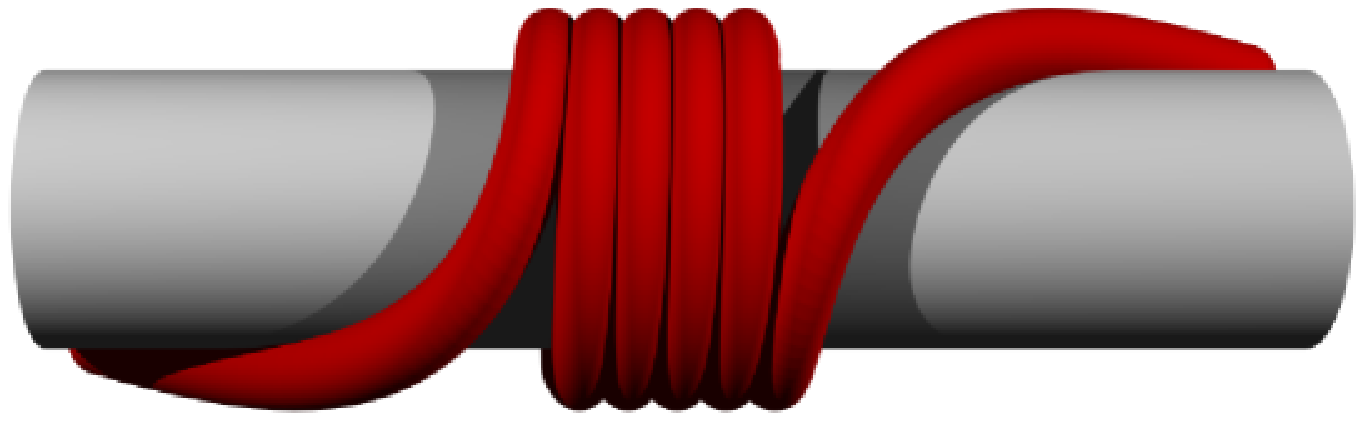}\cr}}$
}
} 
%rechter plaatje           
\rput(0,0.5) {   
\rput(9,2) {
\includegraphics[height=3cm]{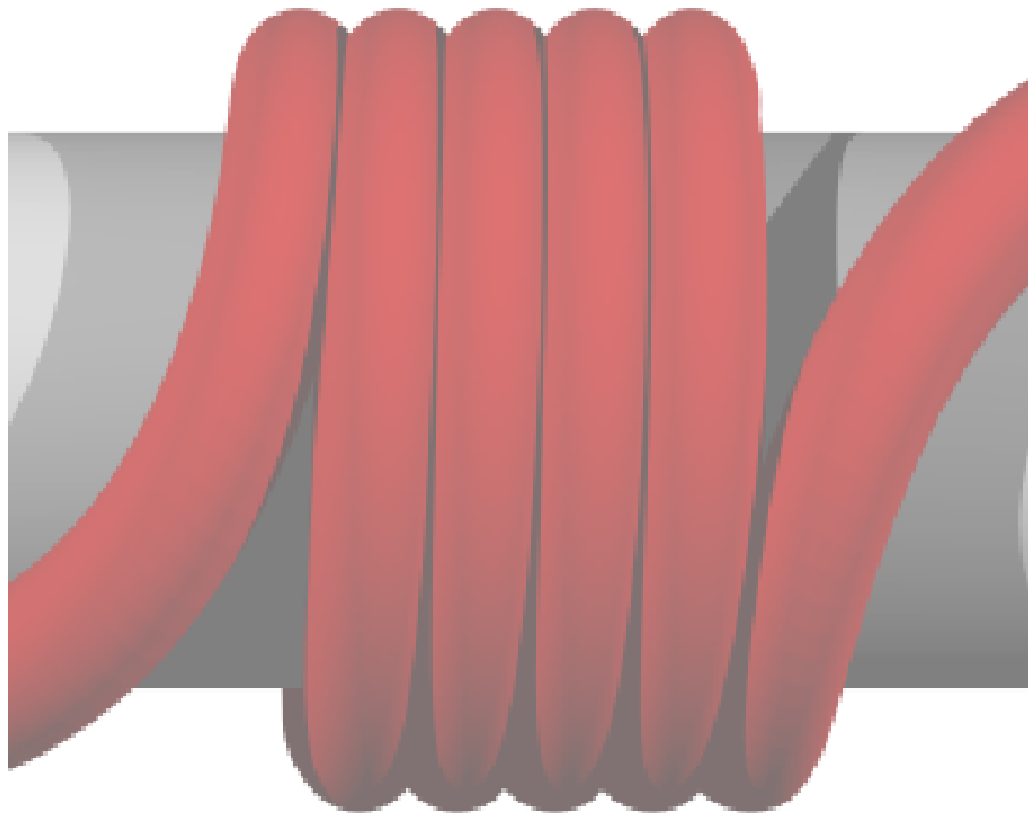}
}
\psset{xunit=0.7cm}
\psset{yunit=0.7cm}
                          
\rput(6.05,0.2) {
                    
\rput(5.7,4){\pscircle*(0,0){.07}}
%\psline{<->}(5.8,4)(6.1,4)
\rput(6.2,4){\pscircle*(0,0){.07}}
%\psline{<->}(6.3,4)(6.6,4)
\rput(6.7,4){\pscircle*(0,0){.07}}
%\psline{<->}(6.8,4)(7.1,4)
\rput(7.2,4){\pscircle*(0,0){.07}}
%\psline{<->}(7.3,4)(7.6,4)
\rput(7.7,4){\pscircle*(0,0){.07}}
                
\rput(-.25,-0.2){
\pspolygon*[linecolor=darkgray](6.1,5)(6.3,5)(6.3,6.6)(6.1,6.6)
\pspolygon*[linecolor=darkgray](6.6,5)(6.8,5)(6.8,6.2)(6.6,6.2)
\pspolygon*[linecolor=darkgray](7.1,5)(7.3,5)(7.3,5.8)(7.1,5.8)
\pspolygon*[linecolor=darkgray](7.6,5)(7.8,5)(7.8,5.4)(7.6,5.4)
}
        
\rput(0,-.5){
\rput(6.1,2){\pscircle*(0,0){.07}}
%\psline{<->}(6.2,2)(6.5,2)
\rput(6.6,2){\pscircle*(0,0){.07}}
%\psline{<->}(6.7,2)(7,2)
\rput(7.1,2){\pscircle*(0,0){.07}}
%\psline{<->}(7.2,2)(7.5,2)
\rput(7.6,2){\pscircle*(0,0){.07}}
%\psline{<->}(7.7,2)(8,2)
\rput(8.1,2){\pscircle*(0,0){.07}}
}
     
\rput(.15,.1){
\pspolygon*[linecolor=darkgray](6,.6)(6.2,.6)(6.2,.2)(6,.2)
\pspolygon*[linecolor=darkgray](6.5,.6)(6.7,.6)(6.7,-0.2)(6.5,-0.2)
\pspolygon*[linecolor=darkgray](7,.6)(7.2,.6)(7.2,-.6)(7,-.6)
\pspolygon*[linecolor=darkgray](7.5,.6)(7.7,.6)(7.7,-1)(7.5,-1)
} 
}
}
\endpspicture
\caption{A typical rod configuration (left; front and back views) that
minimizes energy and satisfies the contact condition. On the right the
bars indicate the contact forces corresponding to the arrows in
Figure~\ref{fig:fg}. (The analysis of this paper assumes zero rod
thickness---in this picture the rod has been fattened for presentation
purposes.)}
\label{fig:typical}
\end{figure}
As mentioned above, the crucial result that the contact set is
connected requires additional assumptions. If we step back from this
rod-on-cylinder model, and allow $a$ and $b$ to be general given
functions, then for a large class of such functions the nonlinear
operator on the left-hand side of~\pref{eq:EL-intro} $N(u)$
satisfies a version of the comparison principle,
\[
Nu_1\geq Nu_2 \quad\Longrightarrow \quad u_1\geq u_2,
\]
(see Definition~\ref{def:compprinc} for the precise statement).  For
such functions $a$ and $b$, any stationary point has a connected
contact set (Theorem~\ref{thm:1}).  The argument is based on the
observation that non-contact in some interval $(\alpha,\beta)$ implies
that $f=0$ on $(\alpha,\beta)$ and therefore that the right-hand side
of~\pref{eq:EL-intro},
\begin{equation}
\label{def:g}
g(x) = \int_{x-1}^x f,
\end{equation}
is non-increasing on $(\alpha,\beta)$ and non-decreasing on
$(\alpha-1,\beta-1)$.

Importantly, however, the functions $a$ and $b$ given in~\pref{def:ab}
are such that the associated operator mostly fails to satisfy this
comparison principle. We therefore also take a different approach, in
which we obtain the same result by only considering global minimizers.
Here we use a different argument, based on constructing other
minimizers by cutting and pasting; the combined condition of
minimization and non-contact in an interval $(\alpha,\beta)$ implies
the existence of additional regions of non-contact outside of the
interval $(\alpha,\beta)$, implying that the right-hand side
of~\pref{eq:EL-intro} is constant on $(\alpha,\beta)$. From this the
result follows (Theorem~\ref{thm:2}).

In both cases, the fact that the contact set is an interval implies
that the boundary of the contact set is `free'---the measure $f$ is
zero on an additional interval of length one extending on both ends of
$\omega_c$.  This implies that the right-hand side $g$ is increasing
and decreasing at the same time---except at points that lie at integer
distance from the two boundary points.  This imposes the specific
structure on $g$ and $f$ that is shown in Figure~\ref{fig:fg}.

\medbreak

The issue of symmetry of minimizers is a subtle one, which again
depends on the presence or absence of a comparison principle.  The
comparison principle simplifies the structure of solutions: all
stationary points are symmetric (up to an unimportant condition on
$b$). Without a comparison principle, and more precisely when
minimization of $F$ favours oscillation, this is no longer true, and
even stationary points that are global minimizers may be asymmetric
(Section~\ref{sec:symmetry}).

\medbreak
Using the characterization of $f$ and $g$ derived earlier we use two
numerical methods to investigate constrained minimizers
(Section~\ref{sec:numerics}); one is a method of direct solution,
using a boundary-value solver, and the other a continuation method.  A
typical solution is shown in Figure \ref{fig:typical}.

\medskip

The simple structure of the functional and the contact condition suggest
that the methods and results of this paper might be applicable to other
systems than this particular rod-on-cylinder model. We therefore
state and prove our results for general functions $a$ and $b$.
The main requirements are
that $a$ and $b$ are smooth and that $a$ is positive; other conditions are
mentioned in the text below.

\section{Problem setting: derivation of the rod-on-cylinder model}
\label{sec:derivation}

%In this section we motivate the choice of the minimization
%problem stated in the introduction. 

\subsection{Kinematics} 

Consider an elastic rod of circular cross-section that is constrained
to lie on a cylinder, and which is subject to a force $T$ and a moment
$M$ at the ends. We assume that at the rod ends, $T$, and $M$ are
maintained parallel to both the axis of the cylinder and the axis of
the rod, but that the loading device leaves the rod ends free to
rotate around the circumference of the cylinder; the ends of the rod
therefore need not be coaxial. The rod is naturally straight and
inextensible, and material cross-sections are assumed to remain
orthogonal to the centerline.  We will derive a minimization problem
for rods of length $2\ell$ and later take the limit $\ell\to\infty$.

In the Cosserat rod theory~\cite[Ch.\ VIII]{antman} the configuration
of this rod is characterized by a right-handed orthogonal rod-centered
coordinate frame of directors, $\{\d1,\d2,\d3\}$, each a function of
the arc-length parameter $s$. The director $\d3$ is assumed parallel
to the centerline tangent, and by the assumption of inextensibility
the centerline curve $\r$ satisfies
\[
\dot \r = \d3,
\]
where the dot denotes differentiation with respect to arc length.  The
strain of the rod is characterized by the vector-valued function $\bm
u$ given by
\[
\bm{{\dot d}}_k = \bm u\times \d k, \qquad k = 1,2,3.
\]
When decomposed as $\bm u = \kappa_1 \d1 + \kappa_2 \d2 + \tau \d3$,
the components may be recognized as the two curvatures $\kappa_{1,2}$
and the twist $\tau$.

We choose a fixed frame of reference $\{\e1,\e2,\e3\}$, where $\e3$ is
parallel to the cylinder axis, and we relate the frame
$\{\d1,\d2,\d3\}$ to this frame by a particular choice of Euler angles
$\{\theta,\psi,\phi\}$~\cite{heijden.98,heijden.01.3}. In this
parametrization $\theta$ is the angle between $\d3$ and~$\e3$ (or
between the centerline and the cylinder axis), $\psi$ characterizes
the rotation around the cylinder axis, and $\phi$ is a partial measure
of the rotation between cross-sections. The condition that the
centerline of the rod lie on the surface of a cylinder of radius $r$
translates into the kinematic condition
\begin{equation}
\label{eq:prevdefpsi}
\dot\psi = \frac1r \sin\theta,
\end{equation}
where the dot denotes differentiation with respect to the arclength
coordinate $s$.  Note that it is natural \emph{not} to restrict $\psi$
to an interval of length $2\pi$. In terms of the remaining degrees of
freedom $\{\ta,\phi\}$ the curvatures and twist are given by
\begin{align*}
\kappa_1 &= \dot\ta \sin\phi - \frac 1r\sin ^2\ta \cos\phi,\\
\kappa_2 &= \dot\ta \cos\phi + \frac 1r\sin ^2\ta \sin\phi,\\
\tau &= \dot \phi + \frac1r \sin\ta \cos\theta.
\end{align*}

\subsection{Energy, work, and a variational problem}
For a given rod the strain energy is given by~\cite{heijden.98},
\begin{align*}
E(\theta,\tau) &= \frac{B}2\int_{-\ell}^\ell(\kappa_1^2 + \kappa_2^2)
+ \frac C2 \int_{-\ell}^\ell \tau^2 \\ &=
\frac{B}2\int_{-\ell}^\ell\dot\ta^2 +
\frac{B}{2r^2}\int_{-\ell}^\ell \sin^4\ta + \frac C2 \int_{-\ell}^\ell
\tau^2. 
\end{align*}
Here $B$ and $C$ are the bending and torsional stiffnesses
respectively. To determine the work done by the tension and moment at
the ends of the rod we need to characterize the generalized
displacements associated with these generalized forces. For the
tension $T$ the associated displacement is the shortening~$S$,
\[ 
S(\ta) = \int_{-\ell}^\ell (1 - \cos \ta).
\] 
The generalized displacement associated with the moment $M$ is the end
rotation, which is well-defined by the assumption of constant end
tangents. It is common to identify the end rotation with a link-like
functional
\[ L 
= \int_{-\ell}^\ell(\dot\phi + \dot\psi) = [\phi + \psi]_{-\ell}^\ell .
\] 
As demonstrated in~\cite{heijden.03}, this identification is correct
in an open set around the undeformed configuration $\ta\equiv0$, but
loses validity when $|\ta|$ takes values larger than $\pi$. Although
nothing we have seen suggests that in an energy-minimizing situation
$\ta$ would take values outside of the admissible interval
$(-\pi,\pi)$, we have no rigorous argument to guarantee that $\ta$
remains inside that interval, and therefore we are forced to assume
this. In terms of the variables $\ta$ and $\tau$ this functional then
takes the form
\[ 
L(\ta,\tau) = 
\int_{-\ell}^\ell\left(\tau + \frac1r \sin\ta(1-\cos\ta) \right). 
\]

Here we assume rigid loading in shortening and dead loading in twist,
i.e.\ we prescribe the shortening $S$ and the moment $M$, which
implies that the tension $T$ and the end rotation $L$ are unknown and
to be determined as part of the solution. This loading condition leads
to the minimization problem
\[
\min \left\{ E(\ta,\tau) - ML(\ta,\tau) : S(\ta) = \sigma \right\}
\]
for given $\sigma>0$. The tension $T$ has a natural interpretation as
a Lagrange multiplier associated with the constraint of $S$.

\medskip

We can simplify this minimization problem by
first minimizing with respect to $\tau$ for fixed
$\ta$, from which we find $\tau\equiv M/C$; re-insertion yields the
final minimization problem
\begin{equation}
\label{pb:min1}
\min \left\{ F(\ta) : S(\ta) = \sigma \right\}
\end{equation}
with
\begin{equation}
\label{def:F}
F(\ta) = \frac{B}2\int\dot\ta^2 + 
\frac{B}{2r^2}\int \sin^4\ta - \frac Mr\int\sin\ta(1-\cos\ta).
\end{equation}
We are interested in localized forms of deformation, in which the
deformation is concentrated on a small part of the rod and in which
boundary effects are to be avoided, and therefore we take an
infinitely long rod and consider $\ta$, $F$, and $S$ to be defined on
the whole real line and assume $\ta\to0$ as $|s|\to\infty$.

\subsection{Behaviour of minimizers}

The Euler-Lagrange equations associated with the minimization
problem~\pref{pb:min1} can be written as a Hamiltonian system with one
degree of freedom,
\begin{equation}
\label{eq:Hamiltonian}
\frac12 \dot\ta^2 + V(\ta) = H,
\end{equation}
for a particular $V$. In this system two independent parameters
remain, which may be interpreted as a scaled cylinder radius $\tilde r
= rM/B$ and a combined loading parameter $m = M/\sqrt{BT}$.

\begin{figure}[!thbp]
\includegraphics[width=11.6cm]{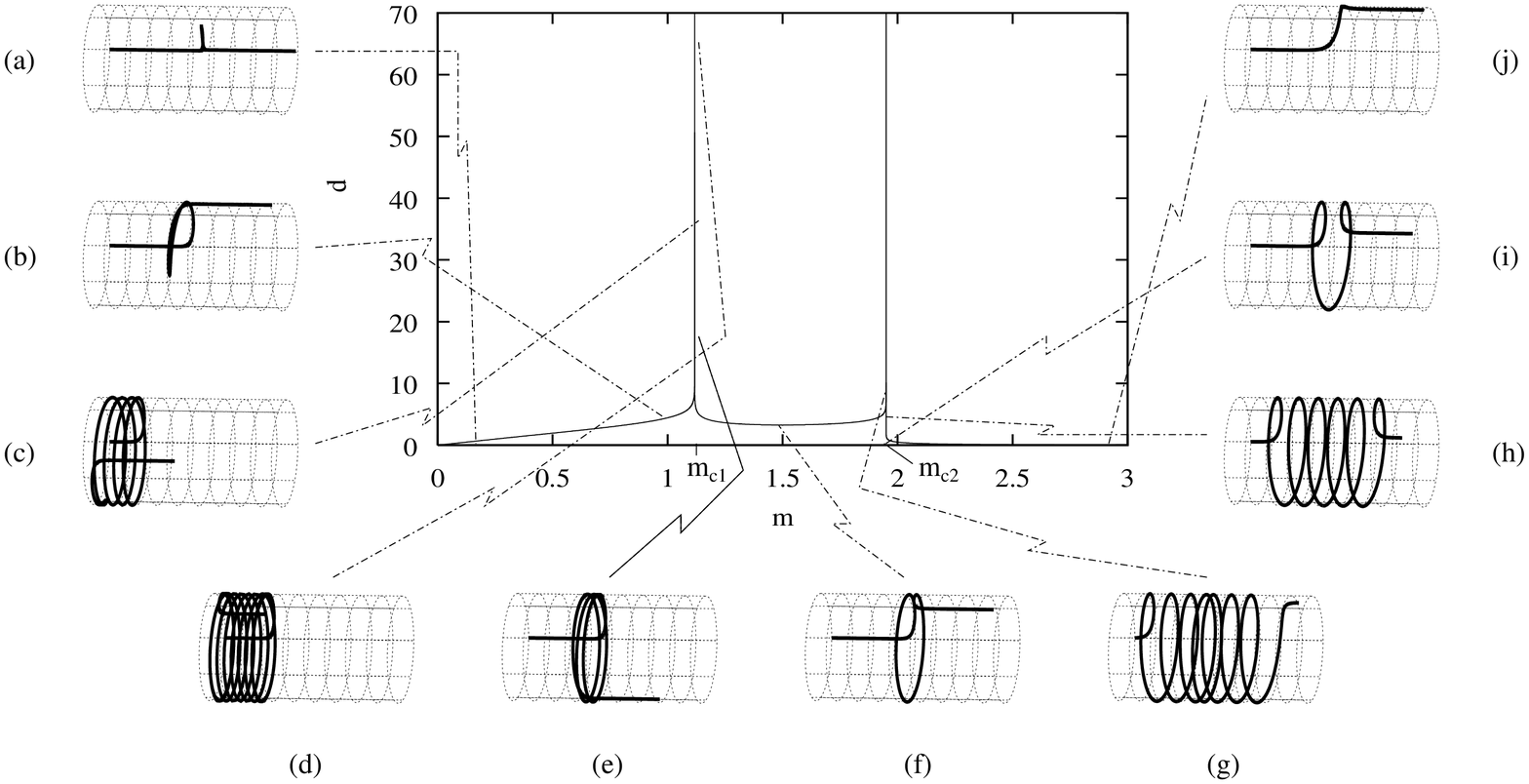}
\caption{A load-displacement diagram showing shortening $d = SM/B$ 
of stationary points as a function of the (combined) load $m$ (from
\cite{heijden.01.3}). Contact effects are not taken into account.  The
peaks divide this diagram into three sections.  The solutions in the
middle section intersect themselves, whilst the solutions on the right
do not.  The section on the left consists of heteroclinic connections
between $\theta=0$ and $\theta=2\pi$ which are not considered here.
For sufficiently large shortening, the rod configuration that has
lowest energy is on the self-penetrating branch, as shown by
Corollary~\ref{cor:contact}.}
\label{fig:gertsoverzicht}
\end{figure}

Solutions of the original minimization problem are orbits of this
Hamiltonian system that are homoclinic to zero, and such orbits have
been studied in detail in~\cite{heijden.01.3}.  Among the findings are
\begin{enumerate}
\item For all values of $\tilde r$ ranges of $m$ exist with orbits
that are homoclinic to the origin;
\item At some parameter points these homoclinic orbits `collide'
with saddle points. The saddle points correspond to helical solutions
(constant angle $\ta$) and close to these collisions the homoclinic
orbit has a large region of near-constant angle $\ta$.
\end{enumerate}
In Figure~\ref{fig:gertsoverzicht} a bifurcation diagram is shown
with two such collisions, one at a forward helix ($0<\ta<\pi/2$, at
$m=m_{c_2}$) and
one at a backward helix ($\pi/2<\ta<\pi$, at $m = m_{c_1}$).

In~\cite{heijden.01.3} the question of stability of these solutions, 
both local and global, was left untouched. 
\drop{The framework 
of~\cite{maddocks.87} % Stability and Folds, ARMA 1987 suggests that
in each `peak' the right-most curve is locally stable; in the case of
two peaks this does not give any argument to distinguish between the
two.  } If we interpret the combined load parameter $m$ as a
(reciprocal) tension $T$ (with the moment $M$ fixed) then the nature
of the bifurcation diagram in Figure \ref{fig:gertsoverzicht},
involving as it does the mechanically conjugate variables $S$ and $T$,
suggests that in each peak the right curve is locally stable
\cite{maddocks.87}. With two peaks occurring however, this does not
allow us to predict where the globally stable solution is located.

In this paper we focus on global energy minimization.
Corollary~\ref{cor:contact} below states that for sufficiently
large shortening, and when contact effects are neglected, 
\emph{global energy minimizers always intersect themselves}. 
It is this result that forms the main motivation of the analysis of
this paper: since energy minimization without appropriate penalization
leads to self-intersection, the non-self-intersection condition is
necessary for physically acceptable solutions.
\subsection{Translation to $(u,\psi)$-coordinates}

To study the case in which self-contact is taken into account, it is
necessary to properly restrict the class of admissible functions in
the minimisation problem~\pref{pb:min1}. In three dimensions a variety
of different descriptions of self-contact exists for rods of finite
thickness, each with subtle advantages and disadvantages (see e.g.\
the introduction of~\cite{gonzalez.01}). For a rod on a cylinder the
situation is simpler, since the freedom of movement is essentially
two-dimensional---similar to that of a curve in a plane. We focus on
rods of zero thickness, and implement non-self-penetration as
non-self-intersection of the centerline. In terms of the unknown
$\ta(\cdot)$ as introduced above, this condition can be written as
\begin{equation}
\label{eq:contact_ta}
z(s_1) - z(s_2) \not= 0
\qquad\text{for all}\qquad
s_1\not=s_2 \text{ with}\quad 
\psi(s_1)-\psi(s_2) = 0 \mod 2\pi,
\end{equation}
where we have used the previous equation \pref{eq:prevdefpsi} for $\psi$ and
the axial coordinate $z$:
\[
\dot\psi = \frac 1r \sin\ta, \qquad \dot z = \cos\ta.
\]
We now make the assumption that $z$ can be written as a function of
$\psi$, or, equivalently, that $\psi$ is monotonic along the rod.
This assumption is satisfied for solutions of the problem without
contact having $\ta < \pi$, as given by
equation~\pref{eq:Hamiltonian}. If we include a contact condition of
the form~\pref{eq:contact_ta}, then we are unable to prove that $\psi$
is monotonic, and in fact it is conceivable that this monotonicity is
only valid for \emph{global} energy minimizers.

\bigskip

Under the assumption that $z$ can be written as a function of $\psi$,
we introduce a dimensionless axial coordinate $\z = z/r$, and write
$'$ for differentiation with respect to $\psi$. The functional $F$
in~\pref{def:F} then transforms to
%\begin{equation}
%\label{eq:Fz}
\[
F(\z) = \frac{Br}{2}\int_0^T \frac{{\z''}^2}{(1 + {\z'}^2)^{\frac 52}} 
+ \frac{B}{2r}\int_0^T \frac{1}{(1 + {\z'}^2)^\frac 32} 
- M \int_0^T \frac{\sqrt{1+{\z'}^2}-\z'}{\sqrt{1+{\z'}^2}},
%\end{equation}
\]
with shortening
\[
S(\z) = r\int_0^T\Bigl[\sqrt{1+{\z'}^2} - \z'\Bigr].
\]
Here $[0,T]$, the domain of definition of $\psi$, is \emph{a priori}
unknown, since the ends of the rod are free to move around the
cylinder.

In these variables non-self-intersection is easily
characterized. Since $\psi$ is monotonic, let us assume it to be
increasing (this amounts to an assumption on the sign of the applied
moment $M$). Admissible functions are defined by the following
condition:
\begin{equation}
\label{cond:contact_psi}
\forall \psi\in [0,T-2\pi]: \z(\psi+2\pi) - \z(\psi) \geq 0.
\end{equation}
Note that it is only necessary to rule out self-intersection 
after a \emph{single}
turn; if contact exists after multiple turns, contact also exists
(potentially elsewhere) after a single turn.

The contact condition~\pref{cond:contact_psi} is the novel part in
this variational problem. In this paper we focus on the effect that
this condition has on the minimization problem, and therefore simplify
by
\begin{itemize}
\item fixing the domain size $T$, and accordingly removing the shortening
constraint;
\item replacing the mechanically correct boundary conditons $\z'=\infty$
by a more convenient condition $\z'=1$.
\end{itemize}
In terms of the new variables $x=\psi/2\pi$ and $u(x) = \z'(\psi)$ we
recover the problem of the introduction.

These boundary conditions can be described as follows.  By prescribing
$\zeta' = u = 1$ at the ends of the rod we fix the angle between the
rod and the centerline to $\pi/4$. By removing the shortening
constraint we allow the ends of the rod to move freely in the axial
direction; in contrast, the fixing of $T$ prevents the rod ends from
moving tangentially. We believe that these changes have little effect
on that part of the rod that is implicated in the contact problem; but
this is a topic of current research.

\subsection{Zero thickness}
\label{subsec:thickness}

\begin{figure}[ht]
\centerline{\psfig{figure=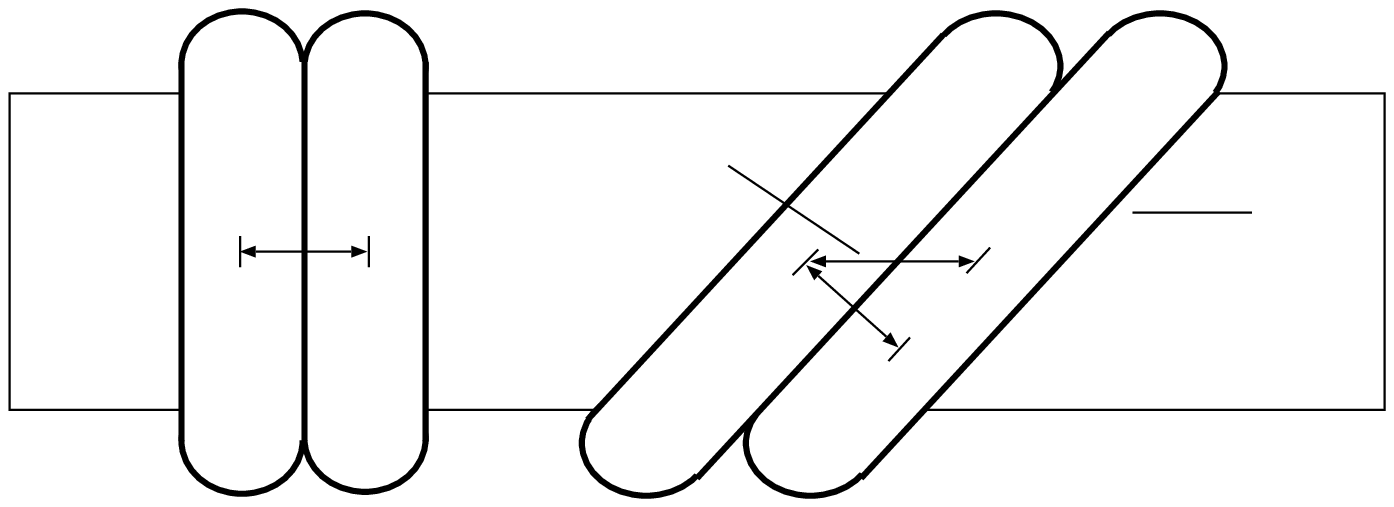,width=9cm}}
\caption{Two configurations of a rod of thickeness $\epsilon$. This
illustrates that for rods with positive thickness one cannot simply
replace the contact condition $Bu \ge 0$ by $Bu \ge \epsilon$; a more
involved condition is necessary.}
\label{fig:fudge}
\end{figure}

The assumption of zero rod thickness can not be relaxed without
introducing important changes in the formulation (see
Figure~\ref{fig:fudge}).  At thickness $\epsilon$, the distance in the
$\zeta$-direction between two parallel consecutive centerlines in
contact is $\epsilon/\sin\theta$, where $\theta$ is the angle between
the centerlines and the cylinder axis.  Therefore non-zero thickness
can not be introduced by simply replacing the right-hand side
in~\pref{cond:contact_psi} by $\epsilon$; the angle of the centerlines
is to be taken into account, implying that the right-hand side
of~\pref{cond:contact_psi} will depend on~$\zeta'$.

To make matters worse, when the centerlines are not parallel,
i.e. when $u=\zeta'$ is not constant, the minimal-distance connection
between two consecutive turns depends on values of $\zeta'$ nearby
(see~\cite{neukirch.02} for a thorough treatment of the geometry
of this issue); it is not clear whether for the present case of a rod
on a cylinder any simpler impenetrability condition can be
found than the well-known global curvature
condition~\cite{gonzalez.01}.

\section{Existence and the contact condition}
\label{sec:existence}

In this section we state precisely the problem under discussion and
show that minimizers exist. We also study the minimization problem
\emph{without} the contact constraint, and show that minimizers will
intersect themselves.

\medskip

Let $U = 1+X$, where $X = H^1_0(0,T)$, and $Y=C([0,T-1])$. 
Let  the functional $F: U \to \R$ be defined as in the Introduction,
\[
F(u) = \int_0^T [\,a(u){u'}^2 + b(u)\,],
\]
and introduce the constraint operator $B:U \to Y$ given by
\[
B(u)(x) = \int_x^{x+1} u.
\]
With the set of admissible functions given as
\[
K := \{u \in U :\ B(u)(x) \ge 0\ \forall x \in [0,T-1]\}
\]
the central problem is
\begin{quote}
{\bf Problem (A)}: Find a function $u^* \in U$
such that
\[
F(u^*) = \min \{ F(u) :\ u \in K\}.
\]
\end{quote}

\bigskip
We first prove existence of minimizers for Problem (A). 
\begin{lemma}
Let $T>0$. Assume that $a(u) \ge a_0 > 0$, and that $b(u)$
is Lipschitz continuous. Then there exists $u^* \in K$ such that
\[
F(u^*) = \min\{\,F(u)\ :\ u \in K\,\}.
\]
\end{lemma}

\begin{proof}
Let $\{u_n\} \subset K \subset U$ be a minimizing sequence.
We first prove that $\int b(u_n)$ is bounded from below.

Since minimization of $F$ is equivalent to minimization of $F-Tb(0)$,
we can assume without loss of generality that $b(0)=0$.  Using the
Lipschitz continuity of $b$ and the Poincar\'e inequality we have
\[
\Modd{b(u_n)}_{L^1} \le c \|u_n\|_{L^1} 
  \leq c_1(T+\Modd{u_n-1}_{L^1}) \leq c (1+\Modd{u_n'}_{L^2}).
\]
Here and below $c$ is a possibly changing constant that does not depend on $n$.
Then
\begin{equation*}
\begin{split}
\int b(u_n) &\ge -c(1+\|u_n'\|_{L^2})\\
        &\ge -c \left(1 + \frac1{a_0} \Big(\int
        a(u_n){u_n'}^2\Big)^{\frac 12}\right)\\ &\ge -c\left(1+
        \frac1{a_0}\Big(F(u_n) -\int b(u_n)\Big)^{\frac 12}\right)\\
        &\ge -c\left(1+ \frac1{a_0}\Big(c - \int b(u_n)\Big)^{\frac
        12}\right).
\end{split}
\end{equation*}
Hence
\begin{equation}
\label{eq:D}
\int b(u_n) \ge -D
\end{equation}
for a suitable constant $D$.

Together with the boundedness of $F(u_n)$, (\ref{eq:D}) implies that
$u_n$ is bounded in~$X$. Hence $\{u_n\}$ contains a subsequence
$\{u_{n_m}\}$ that converges weakly in $X$ to a limit~$u^*$. Since $F$
is lower semicontinuous with respect to weak convergence,
\[
F(u^*) \le \liminf_{m \to \infty} F(u_{n_m}),
\]
implying that $u^*$ is a minimizer.
\end{proof}

As we mentioned in Section~\ref{sec:derivation}, if contact is not
taken into account---if $F$ is minimized in $U$ rather than in the
smaller set $K$---% then minimizers will violate the contact
condition. In the theorem below we actually prove a stronger
statement. We write $F_T$ instead of $F$ to indicate explicitly the
dependence on the interval $[0,T]$.

\begin{theorem}[Minimization without contact condition]
\label{thm:minimizing_rods_intersect}
Assume that $a$ and $b$ are of class $C^1$, and that $a$ is strictly positive. 
Assume that some $\bar u< 1$ exists such that
\begin{equation}
\label{cond:b}
-\infty < \inf_{\R} b < \inf_{u\geq\bar u} b. 
\end{equation}
For each $T>0$, let $u_T$ be
a minimizer corresponding to the minimization problem on domain $[0,T]$,
\begin{equation}
\label{pb:min_unconstrained}
\min \{F_T(u)\ :\ u \in U\}.
\end{equation}
Then  there exists $c>0$ independent of $T$ such that 
\[
\modd{\{x\in[0,T]: u_T(x)\geq  \bar u\}} \leq c\,\bigl(1+\sqrt T\bigr).
\]
\end{theorem}

The function $b$ given in~\pref{def:ab} achieves its minimum at
$u=-\infty$, regardless of the value of $Mr/B$; therefore it satisfies
the condition~\pref{cond:b} for every $\bar u < 1$.

\begin{corollary}[Minimizers violate the contact condition]
\label{cor:contact}
In addition to the conditions of 
Theorem~\ref{thm:minimizing_rods_intersect},
assume that $\bar u < 0$.
If $T$ is sufficiently large, then $B(u_T)(x) < 0$ for some $x \in [0,T-1]$.
\end{corollary}

\begin{proof}[Proof of Theorem \ref{thm:minimizing_rods_intersect}]
We first use a standard argument to give an upper bound on the energy $F_T(u_T)$.
Choose a $T$-dependent constant $\underline u_T<\bar u$ such that 
\[
b(\underline u_T) < \inf_{u\geq\bar u} b(u) 
\qquad \text{and}\qquad
0<b(\underline u_T) - \inf_\R b  \leq T^{-1/2}.
\]
For any $T\in \R^+$ we construct a new continuous symmetric function
$\tilde u _T \in U$ such that $\tilde u_T = \underline u_T$ on
$[1,T-1]$, and $F_T(\tilde u_T) \leq C + Tb(\underline u)$, where $C$
does not depend on $T$. Since $u_T$ minimizes $F_T$, it also follows
that
\begin{equation}
\begin{split}
\label{eq:est2}
F_T(u_T) \le F_T(\tilde u_T) \leq C + Tb(\underline u_T).
\end{split}
\end{equation}
Among other things this inequality implies that for large $T$ a
minimizer $u_T$ can not be the constant function $1$.

The Euler-Lagrange equation associated with this minimization problem
is
\begin{equation}
\label{eq:ELHam}
-2a(u)u'' -a'(u){u'}^2 + b'(u) = 0,
\end{equation}
which can also be written als a one-degree-of-freedom Hamiltonian system
\begin{equation}
\label{eq:ELHam2}
-a(u){u'}^2 + b(u) = H.
\end{equation}
It follows that for any minimizer $u$,
\begin{enumerate}
\item $b(u(x)) = H$ at any stationary point $x$ of $u$;
\item $b(u(x))\geq H$ for all $x\in[0,T]$;
\item $b(1) > H$.
\end{enumerate}
The third statement follows from noting that if $b(1)=H$ then $u\equiv
1$ would be the unique solution of~\pref{eq:ELHam2}.

We now show that any minimizer $u$ is bi-monotonic, i.e.\ increasing
or decreasing away from a minimum or maximum.  Suppose instead that
$u$ has two internal stationary points, a minimum at $x_1$ and a
maximum at $x_2$; assume for definiteness that $0<x_1<x_2<T$. Note
that $u(x_1) < 1 < u(x_2)$, since the solution of the Hamiltonian
system is a periodic orbit oscillating between the values $u(x_1)$ and
$u(x_2)$; the inequality $u(x_1) < 1 < u(x_2)$ follows from the
boundary condition. Now pick a point $x_{12}\in(x_1,x_2)$ such that
$u(x_{12})=1$.

Construct a new function
\[
\tilde u(x) =\begin{cases}
u(x) & 0\leq x\leq x_1\\
u(x_1) & x_1\leq x \leq x_1+T-x_{12} \\
u(x-T+x_{12}) & x_1 + T - x_{12} \leq x \leq T
\end{cases}
\]
Then
\begin{align*}
F_T(\tilde u) &= \int_0^{x_1} [\,a(u) {u'}^2 + b(u)\,]
  + \int_{x_1}^{x_1+T-x_{12}} b(u(x_1))
  + \int_{x_1}^{x_{12}} [\, a(u){u'}^2 + b(u)\,] \\
&= \int_0^{x_{12}} [\,a(u) {u'}^2 + b(u)\,] + \int_{x_1}^{x_1+T-x_{12}} b(u(x_1))\\
&=  \int_0^{x_{12}} [\,a(u) {u'}^2 + b(u)\,] + H(T-x_{12})\\
&< F_T(u).
\end{align*}
Therefore the assumption of two stationary points is contradicted.
Note that by~\pref{eq:ELHam2} the solution also is symmetric in $[0,T]$.

We now return to the sequence of functions $u_T$.
Setting $A = \{x\in[0,T]: u_T(x)\geq  \bar u\}$ we have
\begin{align*}
C+Tb(\underline u_T) &\geq F_T(u_T) \\
  &\geq \modd{A} \inf_{u\geq \bar u}b(u) + (T-\modd{A})\inf_{\R} b \\
  &= \modd{A} \Bigl(\inf_{u\geq \bar u}b(u) - \inf_{\R} b\Bigr) + T\inf_{\R} b,
\end{align*}
so that
\[
\modd{A} \Bigl(\inf_{u\geq \bar u}b(u) - \inf_{\R} b\Bigr) 
  \leq C + T \bigl(b(\underline u_T) - \inf_{\R} b\bigr) \leq c\,(1+\sqrt T).
\]
This concludes the proof.
\end{proof}

\begin{remark}
By a very similar argument one may show the following statement: if
$\min_\R b$ is uniquely achieved at some $\bar u\in\R$, then
\[
\Modd{u_T- \bar u}_{L^\infty\left(\sqrt T,T-\sqrt T\right)} \pijl 0\ 
\qquad\text{as}\qquad T \pijl
\infty.
\]
\end{remark}

\section{The Euler-Lagrange equation}
\label{sec:ELs}
We characterize the duality $(X,X')$ by identifying the smooth
functions on $[0,T]$ with a dense subset of $X'$ via the duality
pairing
\[
\Dual {X'}\xi x X = \int_0^T \xi x.
\]
Similarly we identify $Y'$ with the space of Radon measures
$RM([0,T-1])$ via the same duality pairing, defined for smooth functions,
\[
\Dual {Y'}\eta y Y = \int_0^T \eta y.
\]
Where necessary, we extend Radon measures in $Y'$ by zero outside
of their domain $[0,T-1]$.

\begin{theorem}
\label{thm:EL-equations}
Assume that $a$ and $b$ are globally Lipschitz continuous, and that
$a\geq a_0>0$.  Let $u \in U$ be a solution of Problem (A). Then $u\in
W^{2,\infty}(0,T)$ and there exists a Radon measure $f \in Y'$ such
that
\begin{equation}
\label{eq:EL}
-2a(u(x))u''(x) - a'(u(x)){u'}^2(x) + b'(u(x)) = \int_{x-1}^x f(s)\,ds 
\end{equation}
for almost every $x\in(0,T)$. Moreover
$f \ge 0$ and $\mathrm{supp}\ f  \subset \omega_c$.
\end{theorem}

\begin{definition}
A function $u\in U$ is called a {\it stationary point} if it there exists
a Radon measure $f \in Y'$, with  $f \ge 0$ and $\supp f \subset \omega_c$,
such that~\pref{eq:EL} is satisfied.
\end{definition}

\noindent In the rest of the paper we will often drop the arguments 
in~\pref{eq:EL} and write
\[
-2a(u)u'' - a'(u){u'}^2 + b'(u) = \int_{x-1}^x f.
\]

The proof of Theorem~\ref{thm:EL-equations} follows along the lines
of~\cite{blom.02}.  We fix the function $u$, with contact set
$\omega_c$ defined in~\pref{def:omega_c}, and introduce the cone of
admissible perturbations~$V$,
\[
\V := \{v \in X\ :\ \exists \{\ep _n\}_{n \in \N} \subset \R^+,
\ep_n \to 0\ \text{ such that }\ B(u + \ep _nv) \ge 0 \ \forall
n \in \N\}. 
\]

\begin{lemma}
\label{lem:EL2} Let $u$ be a minimizer. Then
$F'(u) \cdot v \ge 0$ for all $v \in \overline{\V}$.
\end{lemma}

\begin{proof} For any $v \in \V$, the fact $u$ is a minimizer implies that 
\[
F(u + \ep_n v) - F(u) \ge 0 \qquad \text{for all $n \in \N$.}
\]
The conditions on $a$ and $b$ imply that $F$ is Fr\'echet
differentiable in $u$ (this follows from the conditions on $a$ and
inspection of~\pref{eq:Fprime} below), so that
\[
0 \le F(u + \ep_n v) - F(u) = \ep_n F'(u)\cdot v + o(\ep_n||v||_X),
\]
from which it follows that $F'(u) \cdot v \ge 0$. Now, given any $v\in
\overline{\V}$, take a sequence $v_m \subset \V$ that converges to $v$ in $X$.
Since $F'(u):X\to\R$ is a continuous linear operator,
$F'(u)\cdot v_m \to F'(u)\cdot v$. Hence $F'(u)\cdot v\ge 0$ for any
$v \in \overline{\V}$. 
\end{proof}

$\overline V$ can be characterized in a more convenient way:
\begin{lemma}
\label{lem:EL2a}
For any $u \in K$,
\[
\overline{V} = W := \{ v \in X\ :\ Bv \ge 0 \text{ on } \omega_c\}.
\]
\end{lemma}
\noindent
We postpone the proof to the end of this section.

\medskip

$\overline{V}$ is a closed convex cone, with dual cone
\[
\overline{V}^\perp = \{\gamma \in X' :\ \langle \gamma, v\rangle \ge 0\quad
\forall v\in \overline{V}\}.
\]
Let
\[
P = \{ y \in Y :\ y \ge 0\ \mathrm{on}\ \omega_c\}.
\]
This also is a closed convex cone, with dual cone
\[
P^\perp = \{ f \in Y' :\ \langle f, y\rangle \ge 0\quad \forall y \in P\}.
\]

\begin{lemma}
\label{lem:EL1}
If $f \in P^\perp$, then $\supp f \subset \omega_c$ and $f \ge 0$. 
\end{lemma}

\begin{proof}
Given any $y$ with support in $\omega^c_c$, $y \in P$ and $-y \in
P$. Hence $\langle f, y \rangle = 0$ and therefore $\mathrm{supp}\ f
\subset \omega_c$.  Now take $y \in Y$ positive. Then in particular $y
\ge 0$ on $\omega_c$, and $y \in P$. By definition of $P^\perp$ this
implies $f \ge 0$.
\end{proof}

We now use the following Lemma to characterize $\overline{V}^\perp$ in a
different way.

\begin{lemma}
\label{lem:EL3}
Let $Y$ be a Banach space, and $P \subset Y$ a closed convex cone with
dual cone $P^\perp$. Let $X$ be a second Banach space, and $A: X \to Y$ a
bounded linear operator. Let $K$ be the following cone in $X$:
\[
K = \{u \in X :\ Au \in P\}.
\]
Then the dual cone $K^\perp$ can be characterized by
\[
K^\perp = \{A^Tg \in X' : g \in P^\perp\}.
\]
\end{lemma}

The proof of this Lemma can be found in \cite{blom.02}. An immediate
consequence of Lemma \ref{lem:EL3} is
\begin{corollary}
\label{cor:dualcones}
\[
\overline{V}^\perp = \{ B^T f \in X' : f \in P^\perp\}.
\]
\end{corollary}

We now turn to the proof of the main theorem of this section.

\begin{proof}[Proof of Theorem \ref{thm:EL-equations}]
We have seen that, since $u$ is a minimizer, $F'(u) \in
\overline{V}^\perp$ and
\[
\overline{V} = \{v \in X\ :\ B(v) \ge 0 \text{ on } \omega_c\},
\] 
by Lemmas \ref{lem:EL2} and~\ref{lem:EL2a}.  By Corollary
\ref{cor:dualcones} there exists an $f \in P^\perp$ such that $F'(u) =
B^Tf$, and by Lemma \ref{lem:EL1} $\supp f \subset \omega_c$ and $f
\ge 0$.  The conjugate operator $B^T$ is easily seen to be given by
\begin{equation}
\label{def:BT}
B^T\phi(x) = \int_{x-1}^x \phi(s)\, ds
\end{equation}
for a smooth function $\phi\in Y'$, where $\phi$ is implicitly
extended by zero outside of the interval $[0,T-1]$.  We use the same
notation for a general Radon measure $f\in Y'$.

Lastly, direct computation gives
\begin{equation}
\label{eq:Fprime}
F'(u)\cdot v 
        = \int_0^T \bigl[\, 2a(u)u'v' + a'(u){u'}^2v + b'(u)v\, \bigr],
\end{equation}
and hence we obtain the equation
\begin{equation}
\label{eq:ELdistr}
-2[a(u(x))u'(x)]' +a'(u(x)){u'}(x)^2 +b'(u(x)) = \int_{x-1}^x f
\end{equation}
in the sense of distributions. 

We now turn to the statement of regularity. Since $f \in RM([0,T-1])$,
the function $g$ (defined in~\pref{def:g}) is uniformly bounded.
Since all terms in~\pref{eq:ELdistr} except the first are in
$L^1(0,T)$, we have $a(u)u'\in W^{1,1}$, and the lower bound on $a$
implies that $u\in W^{2,1}(0,T)$. Since $W^{2,1}\subset W^{1,\infty}$,
the second term is now known to be in $L^\infty$, and again the lower
bound on $a$ is used to obtain $u\in W^{2,\infty}(0,T)$.  This
regularity of $u$ implies that the distributional
equation~\pref{eq:ELdistr} is also satisfied almost everywhere.
\end{proof}

\medskip

We still owe the reader the proof of Lemma~\ref{lem:EL2a}.
\begin{proof}[Proof of Lemma~\ref{lem:EL2a}]
$\overline{V} \subset W$: Since $B:X\to Y$ is continuous, $W$ is
closed, and therefore it suffices to show that $V \subset W$. Take any
$v \in V$ and $x\in \omega_c$. Then $B(u + \ep_n v)(x) \ge0$, and by
definition of $\omega_c$, $Bu(x) = 0$, implying that $Bv(x) \ge 0$.
It follows that $v \in W$.

$W \subset \overline V$: First consider $w \in W$ such that $\supp
(Bw)_-$ (the support of the negative part of $Bw$) is contained in
$\omega_c^c$. We claim that $w \in V$, for which we have to show that
there exists
\[
\{\ep _n\} \subset \R^+,\ \ep_n \to 0,
\text{ such that } B(u + \ep _n w) \ge 0\ \forall n \in \N.
\] 

For $x \in \omega_c$, $Bu(x) = 0$, and since $Bw(x) \ge 0$ we have
$B(u + \ep_n w)(x) \ge 0$ for any sequence $\{\ep_n\} \subset \R^+$.
%So let $x
%\in \omega_c^c$. 
For the complement $\omega_c^c$, note that since $\supp (Bw)_-$ is
compact and contained in the open set $\omega_c^c$, there exists
$\delta > 0$ such that $Bu \ge \delta > 0$ on $\supp (Bw)_-$. Hence,
if $\ep_n \le \delta\Modd{Bw}_{L^\infty}^{-1}$, then $Bu + \ep _nBw
\ge 0$ on $\supp (Bw)_-$. Note that $Bu \ge 0$ on $[0,T-1]$, and $Bw
\geq 0$ on $(\supp (Bw)_-)^c$. This means that $Bu + \ep_n Bw \ge 0$
on $\omega_c^c$. Together with $Bw \ge 0$ on $\omega_c$, this implies
$w \in V$.

Finally, consider a general $w \in W$. Fix a smooth function $\phi \in
X$ with $\phi > 0$ on $(0,T)$; note that $B\phi \geq c > 0$. We
approximate $w$ by the function $w_{\ep} := w + \ep \phi$. We claim
that $\supp (Bw_{\ep})_- \subset \omega_c^c$ for sufficiently
small~$\ep > 0$. It then follows that $w_{\ep} \in V$ and $w_{\ep} \to
w$, implying that $w\in \overline{V}$.

To prove the claim, note that $w \in X \subset L^{\infty}$. Hence $Bw$
is Lipschitz continuous, with Lipschitz 
constant $2||w||_{L^{\infty}}$. Hence, for
small enough $\ep$,
\begin{equation}
\label{eq:w_eps}
\begin{split}
Bw_{\ep}(x) &= Bw(x) + \ep B\phi(x) \\
        &\ge  Bw(y) +\ep B\phi(y) - 3||w||_{L^{\infty}}|x - y|.
\end{split}
\end{equation}
Suppose $Bw_{\ep}(x) < 0$ and $y \in \omega_c$. Then $Bw(y) \ge 0$, and 
by (\ref{eq:w_eps}),
\[
-\ep B\phi(y) > -3||w||_{L^{\infty}}|x - y|,
\]
or
\[
|x - y| > \frac{\ep B\phi(y)}{3||w||_{L^{\infty}}}.
\]
Therefore $d(\supp(Bw)_-,\omega_c) > C\ep$
%So if $x \in \supp (Bw_{\ep})_-$ and $y \in \omega_c$, then
%$d(x,y) > C\ep$ 
for a suitable $C > 0$. Hence $\supp
(Bw_{\ep})_-$ $\subset$ $\omega_c^c$ for small enough $\ep$, which
proves the claim. 
\end{proof}

\section{Characterization of stationary points}
\label{sec:characterization}

For this section we assume that the conditions of
Theorem~\ref{thm:EL-equations} are met.

\begin{lemma}
\label{lem:ctct-periodicity}
Let $u$ be a stationary point, and let $g$ be defined as
in~\pref{def:g}.
\begin{enumerate}
\item For all $x\in \omega_c$, $u(x) = u(x+1)$ and $u'(x) \le u'(x+1)$. 
\item If $\omega_c$ contains an interval
$[x_0, x_1]$, then 
\begin{itemize}
\item $u'(x) = u'(x+1)$ for all $x \in (x_0, x_1)$;
\item $u''(x) = u''(x+1)$  and $g(x) = g(x+1)$ for almost all $x\in(x_0,x_1)$. 
\end{itemize}
\end{enumerate}
\end{lemma}

This lemma imposes an interesting form of periodicity on the solution
and the right-hand side $g$.  Although the constraint is a non-local
one, on an interval of contact of length~$L$ the solution actually
only has the degrees of freedom of an interval of length one; the
other values follow from this assertion.

\begin{proof}
Since $x \in \omega_c$, 
\[
\int_{x}^{x+1} u = 0.
\]
Hence, since $Bu(x) = \int_x^{x+1}u \in W^{3,\infty}$, and $Bu \ge 0$,
\[
0 = \frac d{dx}\int_{x}^{x+1} u = u(x+1) - u(x),
\]
and
\[
0 \le \frac{d^2}{dx^2} \int_{x}^{x+1} u
= u'(x+1) - u'(x).
\]
If $Bu=0$ on $[x_0,x_1]$, then the inequality above becomes an equality
a.e. on the interior $(x_0,x_1)$, implying that
\[
u'(x) = u'(x+1) \text{ on }(x_0, x_1).
\]
The periodicity of $u''$ and $g$ now follow from \pref{eq:EL}.
\end{proof}

\begin{lemma}
\label{lem:int_g}
Let $u$ be a stationary point, and assume that $\omega_c$ contains an
interval~$I$. Then
\[
\int_x^{x+1}g
\]
is constant on $\Int I$.
\end{lemma}

\begin{proof}
By Lemma \ref{lem:ctct-periodicity} $u(x) = u(x+1)$ for all $x \in I$,
and $u''(x) = u''(x+1)$ a.e.\ on~$I$.
In addition, $u'(x) = u'(x+1)$ for all $x \in \Int{I}$. Hence
\begin{equation}
\label{eq:1}
x \mapsto \int_x^{x+1} \bigl[-2a(u)u'' -a'(u){u'}^2 + b'(u)\bigr] 
\end{equation}
is constant on $I$. But by~\pref{eq:EL}, (\ref{eq:1}) is
equal to 
\[
\int_x^{x+1} \int_{s-1}^s f = \int_x^{x+1} g.
\]
\end{proof}

\medbreak

The following two lemmas and the theorem that follows are essential in
determining the structure of the right-hand side $g$ and therefore of
the measure $f$.  The main argument is the following. The function $g$
has no reason to be monotonic; its derivative in $x$ equals
$f(x)-f(x-1)$, and although $f$ is a positive measure this difference
may be of either sign. However, if for instance a left end point~$x_0$
of~$\omega_c$ is flanked by a non-contact interval $(x_0-1,x_0)$, then
the measure $f$ is zero on that interval, and the function $g$ is
non-decreasing on $(x_0,x_0+1)$. It is this argument, repeated from
both sides, that allows us to determine completely the structure of
the function $g$ and the underlying measure $f$.

\medskip
{\bf Notation} Let $[x_0, x_1] \subset \omega_c$. Define
\begin{equation}
\label{def:p}
p \equiv x_1 - x_0 \modu 1,
\end{equation}
and
\begin{equation}
\label{def:P}
P = \min\{n\in\N: n\geq x_1-x_0\}.
\end{equation}

Throughout the rest of this paper $\tau$ is the translation operator
defined by 
\begin{equation}
\label{def:tau}
(\tau u)(x)= u(x+1).
\end{equation}

\begin{lemma}
\label{structure_g_1}
Let $u$ be a stationary point, such that $\omega_c$ contains an
interval $[x_0, x_1]$. Assume furthermore that
\begin{equation}
\label{ass:disj}
\supp f \cap (x_0 -1, x_0) = \emptyset.
\end{equation}
Then
\begin{enumerate}
\item if $x_1 - x_0 \in \N$, then
$g$ does not decrease on each of the subintervals
\[
(x_0 + i, x_0 + i + 1),\ i = 0, 1, \ldots, P;
\]
\item if $x_1 - x_0 \not\in \N$, then
$g$ does not decrease on each of the subintervals
\[
(x_0 + i, x_0 + i + 1),\ i = 0, 1, \ldots, P-1,
\]
nor does it on
\[
(x_0 + P, x_1 + 1).
\]
\end{enumerate}
\end{lemma}

\begin{proof}
On $(x_0, x_0+1)$,
\[
g' = f - \tau^{-1} f \;\overset{\pref{ass:disj}}=\; f \ge 0,
\]
and therefore $g$ is non-decreasing on $(x_0, x_0+1)$. 
By Lemma \ref{lem:ctct-periodicity}, $g(x) = g(x +1)$ for almost all
$x \in (x_0, x_1)$. This implies that on each consecutive interval
$(x_0 + i, x_0 + i + 1)$, $i = 1, \ldots, P-1$, $g$ does not decrease.
By the same reasoning, if $x_1 - x_0 \in \N$, then this also holds for
$(x_0 + P, x_0 + P + 1) = (x_1, x_1 + 1)$. If not, then it holds
for $(x_0+P, x_1 + 1)$.
\end{proof}

\begin{remark}
Let $u$ be a stationary point. Define the mirror image $v(x) =
u(T-x)$, and $h(x) = f(T- x - 1)$. Then $(v, h)$ solves
\[
\left\{
\begin{array}{l}
-2a(v)v'' - a'(v){v'}^2 + b'(v)= \int_{x-1}^x h,\\[2mm]
v(0) = v(T) = 1,
\end{array}
\right.
\]
and hence is also a stationary point.
\end{remark}
Applying Lemma \ref{structure_g_1} to
$(v, h)$ yields for $(u, f)$:

\begin{lemma}
\label{structure_g_2}
Let $u$ be a stationary point such that $\omega_c$ contains an
interval $[x_0, x_1]$. Assume furthermore that
\[
\supp f \cap (x_1 + 1, x_1 + 2) = \emptyset.
\]
\begin{enumerate}
\item if $x_1 - x_0 \in \N$, then
$g$ does not increase on each of the subintervals
\[
(x_0 + i, x_0 + i + 1),\ i = 0, 1, \ldots, P;
\]
\item if $x_1 - x_0 \not\in \N$, then
$g$ does not increase on each of the subintervals
\[
(x_0 + p + i, x_0 + p + i + 1),\ i = 0, 1, \ldots, P-1,
\]
nor does it on
\[
(x_0, x_0 + p).
\]
\end{enumerate}
\end{lemma}

To combine the previous two Lemmas, let
\begin{equation}
\begin{array}{lll}
\label{eq:XiYi}
X_i&= x_0 +  i,&\ i = 0, \ldots, P,\\
Y_i&= x_0 + p + i,&\ i = 0, \ldots, P.
\end{array}
\end{equation}

\begin{theorem}
\label{thm:structure_g_3}
Let $u$ be a stationary point such that the contact set $\omega_c$
contains an interval $[x_0, x_1]$. Suppose that
\[
\supp f \cap \bigl\{(x_0 - 1, x_0)\ \cup\ (x_1 + 1, x_1 +2)\bigr\} = \emptyset.
\]
Then there exists $G\in\R$ such that 
\begin{enumerate} 
\item if $x_1 - x_0 \in \N$, then $g \equiv G$ on $(x_0, x_1 + 1)$,
and
\[
f|_{(x_0-1, x_1+2)} = G\sum _{i=0}^P \delta(x - X_i).
\]
\item if $x_1 - x_0 \not\in \N$, then 
\begin{equation}
\label{def:g1g2}
g(x) = \left\{
\begin{array}{lll}
g_1 & \text{ on } [X_i, Y_i], & i=0,\ldots,P,\\
g_2 & \text{ on } [Y_i, X_{i+1}],& i=0,\ldots,P-1,
\end{array}
\right.
\end{equation}
and
\[
f|_{(x_0-1, x_1+2)} = \sum _{i=0}^{P-1} a_i\delta(x - X_i) +
b_i\delta(x - Y_i),
\]
where $a_i = (G - \frac iP)g_1$ and $b_i = \frac{G + i}{P}g_1$,
and
\begin{gather}
\label{def:g_1}
g_1 := \frac{GP}{P + 1 - p}\in \Bigl(\frac {PG}{P+1},G\Bigr),\\[2\jot]
g_2 := \frac{G(P+1)}{P + 1 - p} = \frac {P+1}Pg_1 \in \Bigl(G,\frac {(P+1)G}P\Bigr).
\label{def:g_2}
\end{gather}
\end{enumerate}
\end{theorem}

\begin{proof}
$(1)$ $x_1 - x_0 \in \N$.

By Lemma \ref{structure_g_1}, $g$ does not decrease on the intervals
$(X_i, X_{i+1})$, $i = 0, 1, \ldots, P$, and by Lemma
\ref{structure_g_2} $g$ does not increase on these intervals either.
Hence $g$ is constant on each interval. By Lemma \ref{lem:int_g} the
constant is the same on each interval, i.e.\ that $g \equiv G$ on
$(x_0, x_1 + 1)$. This also implies that within the interval $(x_0 -1,
x_1 + 2)$, $f$ can only have support in the points $x_0 = X_0, X_1,
\ldots, X_P = x_1$, yielding the formula for $f$ in the statement of
the theorem.

$(2)$ $x_1 - x_0 \not\in \N$.

Combining Lemma \ref{structure_g_1} and \ref{structure_g_2}, we find
that $g$ is constant on each interval $(X_i, Y_i)$ and
$(Y_i,X_{i+1})$, $i = 0, 1, \ldots, P-1$, and on $(X_P, Y_P)$.  By
Lemma \ref{lem:ctct-periodicity}, $g(x) = g(x+1)$ for almost all $x
\in (x_0, x_1)$, and hence $g$ takes three values, $0$, and $g_1$ and
$g_2$ (say) on $(x_0, x_1 + 1)$. We choose $g = g_1$ on $(X_i,Y_i)$,
$i = 0,1, \ldots, P$, and $g = g_2$ on the intervals inbetween,
$(Y_i,X_{i+1})$, $i = 0, 1, \ldots, P-1$; outside of the interval
$(x_0, x_1)$, $g$ vanishes. By Lemma \ref{lem:int_g},
\begin{equation}
\label{eq:g1g2}
G = \int_{x}^{x+1} g = \int_{x}^{x+p}g_1 +\int_{x + p}^{x+1}g_2 = pg_1
+ (1 - p)g_2.
\end{equation}
Either $g_1 = g_2 = G$ or $g_1 < G < g_2$. The first case implies that
$g \equiv G$ on $[x_0, x_1 + 1]$. This implies that $f$ does not only
have support in $X_0$, $X_1$, $\ldots$, $X_{P-1}$, but by reasoning
for the mirror image $(v, h)$ it also implies that $f$ has support in
$Y_0, Y_1, \ldots, Y_{P-1} =x_1$. This is impossible. Hence $g_1 < G <
g_2$.  The support of $f$ on $(x_0 - 1, x_1 + 2)$ is now seen be to
limited to the set given in the statement of the Theorem.

Thus we conclude that $f$ is a sum of delta functions, but we still
have to determine the weights $a_i$ and $b_i$. Since $f = 0$ on $(x_0
- 1, x_0)$, we have $g_1 = g(x_0+) = f(x_0)$.  Here we abuse notation,
and write $f(x)$ for the weight of the Dirac delta function at
$x$. Now we have the following recurrence relations:
\[
\begin{array}{l}
f(X_i) + f(Y_i) = g_2,\\
f(Y_i) + f(X_{i + 1}) = g_1,
\end{array}
\]
for $i = 0, 1, \ldots, P-1$.  Solving this system we
obtain
\[
\begin{array}{l}
f(X_i) = f(X_0) - i(g_2 - g_1) = g_1 - i(g_2 - g_1),\\
f(Y_i) = (i+1)(g_2 - g_1).
\end{array}
\]
In addition, since $x_1 = Y_{P-1}$, $f(x_1) = P(g_2 - g_1)$.
On the other hand, $g_1 = h(T - x_1 -1) = f(x_1)$. This implies
\[
g_2 = \frac {P+1}P g_1.
\]
To conclude,
\[
a_i = f(X_i) = \big(G - \frac iP\big)g_1,
\]
and
\[
b_i = f(Y_i) = \frac {i + 1}Pg_1.
\]
By Lemma \ref{lem:int_g},
\[
pg_1 + (1 - p)\frac{P+1}Pg_1 = G,
\]
which yields
\[
p = P\Bigl(1 - \frac G{g_1}\Bigr) + 1.
\]
Solving for $g_1$ now yields all required results.
\end{proof}

\begin{figure}[!hbtp]
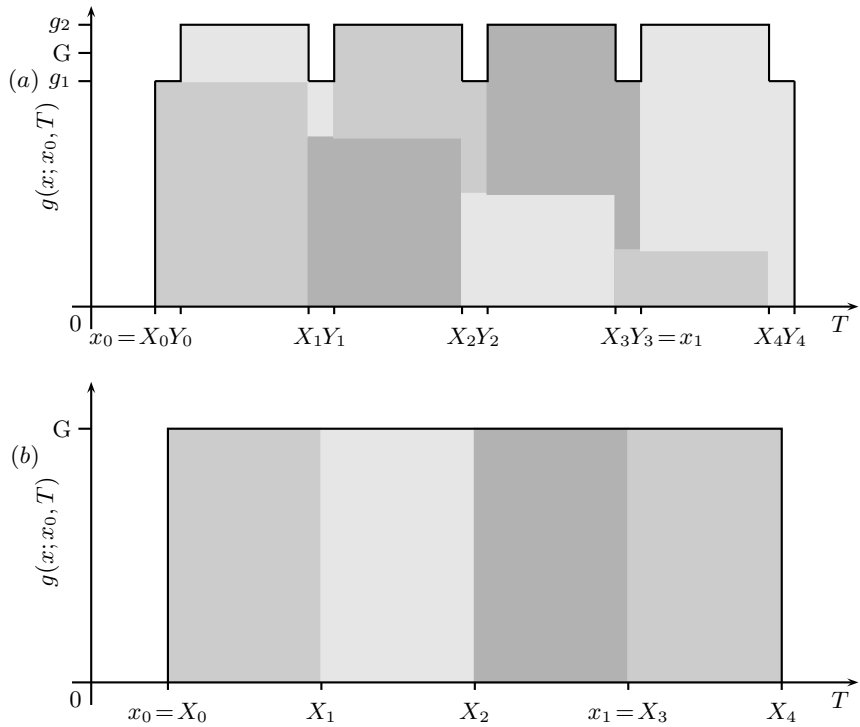

\pspicture(-1,-1)(12,10)
%\psgrid(0,0)(-1,-1)(12,10)

%%%%% (a)  %%%%%%%%%%%%%%

\psset{xunit=.85cm}

\newgray{gray1}{0.9}
\newgray{gray2}{0.8}
\newgray{gray3}{0.7}

\rput(0,-10){
\pspolygon[fillstyle=solid,fillcolor=gray2,linecolor=gray2](1,15)(1,18)
(3.4,18)(3.4,15)
\pspolygon[fillstyle=solid,fillcolor=gray1,linecolor=gray1]
(1.4,18)(1.4,18.75)(3.4,18.75)(3.4,18)
\pspolygon[fillstyle=solid,fillcolor=gray1,linecolor=gray1]
(3.4,17.25)(3.4,18)(3.8,18)(3.8,17.25)
\pspolygon[fillstyle=solid,fillcolor=gray3,linecolor=gray3]
(3.4,15)(3.8,15)(5.8,15)(5.8,17.25)(3.8,17.25)(3.4,17.25)
\pspolygon[fillstyle=solid,fillcolor=gray2,linecolor=gray2]
(5.8,17.25)(3.8,17.25)(3.8,18.75)(5.8,18.75)
(5.8,18)(6.2,18)(6.2,16.5)(5.8,16.5)
\pspolygon[fillstyle=solid,fillcolor=gray1,linecolor=gray1]
(5.8,15)(8.2,15)(8.2,16.5)(5.8,16.5)
\pspolygon[fillstyle=solid,fillcolor=gray3,linecolor=gray3]
(6.2,16.5)(8.2,16.5)(8.2,15)(8.6,15)(8.6,18)
(8.2,18)(8.2,18.75)(6.2,18.75)
\pspolygon[fillstyle=solid,fillcolor=gray2,linecolor=gray2]
(8.2,15)(10.6,15)(10.6,15.75)(8.2,15.75)
\pspolygon[fillstyle=solid,fillcolor=gray1,linecolor=gray1]
(8.6,18.75)(8.6,15.75)(10.6,15.75)(10.6,15)(11,15)
(11,18)(10.6,18)(10.6,18.75)

\psline(1,15)(1,18)
\psline(11,15)(11,18)

\psline(1,18)(1.4,18)(1.4,18.75)(3.4,18.75)(3.4,18)
\rput(2.4,0){\psline(1,18)(1.4,18)(1.4,18.75)(3.4,18.75)(3.4,18)}
\rput(4.8,0){\psline(1,18)(1.4,18)(1.4,18.75)(3.4,18.75)(3.4,18)}
\rput(7.2,0){\psline(1,18)(1.4,18)(1.4,18.75)(3.4,18.75)(3.4,18)}
\psline(10.6,18)(11,18)

\uput[d](0.6,14.9){$x_0\!=\!X_0$}
\uput[d](1.4,14.9){$Y_0$}
\psline(1,15)(1,14.9)
\psline(1.4,15)(1.4,14.9)

\uput[d](3.4,14.9){$X_1$}
\uput[d](3.8,14.9){$Y_1$}
\psline(3.4,15)(3.4,14.9)
\psline(3.8,15)(3.8,14.9)

\uput[d](5.8,14.9){$X_2$}
\uput[d](6.2,14.9){$Y_2$}
\psline(5.8,15)(5.8,14.9)
\psline(6.2,15)(6.2,14.9)

\uput[d](8.2,14.9){$X_3$}
\uput[d](9,14.9){$Y_3\!=\!x_1$}
\psline(8.2,15)(8.2,14.9)
\psline(8.6,15)(8.6,14.9)

\uput[d](10.6,14.9){$X_4$}
\uput[d](11,14.9){$Y_4$}
\psline(10.6,15)(10.6,14.9)
\psline(11,15)(11,14.9)

\uput[dl](0,15){0}
\uput[dl](12,15){$T$}

\uput[l](-.1,18.375){G}
\psline(0,18.375)(-.2,18.375)

\uput[l]{90}(-.3,17){$g(x;x_0,T)$}

\uput[l](-.1,18){$g_1$}
\uput[l](-.1,18.75){$g_2$}

\psline(0,18)(-.2,18)
\psline(0,18.75)(-.2,18.75)

\psline{->}(-.3,15)(12,15)
\psline{->}(0,14.7)(0,19)

\uput[l](-.6,18){$(a)$}
}

%####### subfigure (b). ###########

\rput(0,-15) {
\uput[l]{90}(-.3,17){$g(x;x_0,T)$}

\pspolygon[fillstyle=solid,fillcolor=gray2,linecolor=gray2]
(1.2,15)(1.2,18.375)(3.6,18.375)(3.6,15)
\pspolygon[fillstyle=solid,fillcolor=gray1,linecolor=gray1]
(3.6,18.375)(3.6,15)(6,15)(6,18.375)
\pspolygon[fillstyle=solid,fillcolor=gray3,linecolor=gray3]
(8.4,18.375)(8.4,15)(6,15)(6,18.375)
\pspolygon[fillstyle=solid,fillcolor=gray2,linecolor=gray2]
(8.4,18.375)(8.4,15)(10.8,15)(10.8,18.375)

\psline(1.2,15)(1.2,18.375)(10.8,18.375)(10.8,15)

\uput[d](1.2,14.9){$x_0\!=\!X_0$}
\psline(1.2,15)(1.2,14.9)
\uput[d](3.6,14.9){$X_1$}
\psline(3.6,15)(3.6,14.9)
\uput[d](6,14.9){$X_2$}
\psline(6,15)(6,14.9)
\uput[d](8.4,14.9){$x_1\!=\!X_3$}
\psline(8.4,15)(8.4,14.9)
\uput[d](10.8,14.9){$X_4$}
\psline(10.8,15)(10.8,14.9)

\psline{->}(-.3,15)(12,15)
\psline{->}(0,14.7)(0,19)

\uput[dl](0,15){0}
\uput[dl](12,15){$T$}

\uput[l](-.1,18.375){G}
\psline(0,18.375)(-.2,18.375)

\uput[l](-.6,18){$(b)$}
}
\endpspicture
\label{fig:gs}
\caption{A generic picture of $g(x;x_0,T)$ for $T - x_0 \not\in \N$
$(a)$ and for $T - x_0 \in \N$ $(b)$. The light gray shaded areas
represent the contributions of the individual delta functions of the
corresponding Radon measures $f$. As an example, in $(b)$ $f$ consists
of four Dirac deltas, all with mass $G$, at $x_0 =X_0, x_0 + 1=X_1,
\ldots, x_0 + 3=X_3$.}
\end{figure}

As we will see in the next Section, the contact set of $u$ is
connected in many important cases. Hence Theorem
\ref{thm:structure_g_3} allows us to give concise expressions for~$g$ in
cases that $\omega_c$ is an interval of positive length (using the
Heaviside function~$H$):
\begin{corollary}
If the contact set is an interval of positive length, then
$g$ equals the explicit function
\begin{multline}
\label{eq:g_explicit}
g(x;x_0,x_1,G) =\\ 
\left\{
\begin{array}{lr}
H(x - x_0) - H(x - x_1 - 1) &\mathrm{if\ } x_1 - x_0 \in \N, \\
g_1\big(H(x - x_0) - 
H(x - x_1 - 1)\big) + \\
\quad \quad 
{}+ (g_2 - g_1)\sum_{i=1}^P \bigl[H(x - X_i) - H(x - Y_i)\bigr] &\mathrm{if\ } x_1 -x_0 \notin \N.
\end{array}
\right.
\end{multline}
Here the coefficients $g_{1,2}$ are computed from $x_0$, $x_1$, and $G$ by~\pref{def:p},
\pref{def:P}, \pref{def:g_1}. and~\pref{def:g_2}.
\end{corollary}
Figure~\ref{fig:gs} shows examples of both cases. For the remaining
two cases of a stationary point that has a single or no contact point,
$g$ is immediately clear: with a single contact point,
\[
g(x) = \left\{
\begin{array}{ll}
m & \text{ on } [x_0, x_0+1],\\
0 & \text{ otherwise,}
\end{array}
\right.
\]
for a suitable constant $m\geq0$, while when there is no contact then
obviously $g\equiv 0$.

\drop{
\medskip

A useful result states that the values of $u$ at jump points are
equal:

\begin{lemma}
\label{lem:equal-u}
Under the conditions and notation of Theorem~\ref{thm:structure_g_3}, 
\begin{equation}
\label{eq:equal-u}
u(x_0) = u(x_0+p) = u(x_0+1) = u(x_0+1+p) = \dots = 
u(x_1+1).
\end{equation}
\end{lemma}

\begin{proof}
By multiplying~\pref{eq:EL} with $u'$ and integrating one finds
that the function $H$, defined by
\begin{equation}
\label{def:H}
H := -a(u){u'}^2 + b(u) - gu,
\end{equation}
is piecewise constant, and that $H$ and $g$ jump at the same values of
$x$ (these are the positions occurring in~\pref{eq:equal-u}).  Since
$u(x+1)=u(x)$ for all $x\in[x_0,x_0+p]$ (assuming $p>0$), and since
$g=g_1$ on both $(x_0,x_0+p)$ and $(x_0+1,x_0+1+p)$, the constant
value of $H$ also is the same on these two intervals.  Repeating this
argument for the other intervals and summarizing we find that both $H$
and $g$ take three values (or two if $p=0$): a value $H_0\in\R$ and
$g_0=0$ outside of the extended contact interval $[x_0,x_1+1]$, and
two (or one) values $(H_1,g_1)$ and $(H_2,g_2)$ alternatively on
$[x_0,x_1+1]$.

At any of the \emph{interior} jump points, i.e.\ at all jump points
except $x_0$ and $x_1+1$, we have $[H] = -[g] u$ where $[H] =
\pm(H_2-H_1)$ and $[g] = \pm(g_2-g_1)$.  Regardless of the sign this
equation has only one solution $u$.  For the remaining two points
$x_0$ and $x_1+1$ the result follows from
Lemma~\ref{lem:ctct-periodicity}.
\end{proof}

}

\section{The contact set is an interval}
\label{sec:interval}

In order to extract more information on the right-hand side $g$ and
the measure $f$ than that given by Theorem~\ref{thm:EL-equations} we
study two cases. In the first case we assume that the operator given
by the left-hand side in~\pref{eq:EL} satisfies a version of the
classical comparison principle. In the second case we restrict
ourselves to global minimizers.

\begin{definition}
\label{def:compprinc}
Let $N$ be a (non)linear operator on $U$.
$N$ is said to satisfy the comparison principle if for
any $[x_0, x_1] \subset [0,T]$,
\begin{equation*}
\left.
\begin{array}{ll}
Nu_1 \le  Nu_2,\\
u_1(x_0) \le u_2(x_0),\\ u_1(x_1) \le u_2(x_1),
\end{array}
\right\}
\implies\ u_1 \le u_2 \ \mathrm{on\ }[x_0,x_1].
\end{equation*}
\end{definition}
\noindent 
See e.g.~\cite{gilbarg.77} or \cite{protter.67} for a general exposition. Operators
of the type considered here, i.e.\ 
\[
Nu:= -2a(u)u'' -a'(u){u'}^2 + b'(u),
\]
may fail to satisfy the comparison principle for two reasons. First,
the zero-order term $b'(u)$ need not be increasing in $u$; for
instance, the operator $u \mapsto -u'' - u$ does not satisfy the
comparison principle on any interval of length $2\pi$ or more.  In a
slightly more subtle manner, the prefactor $a(u)$ of the second-order
derivative may also invalidate the comparison principle; see
e.g.~\cite[Section~10.3]{gilbarg.77} for an example.

We conjecture that the `true' rod functions $a$ and $b$ given
in~\pref{def:ab} do not give rise to a comparison principle: $b'$ is
not monotonic, suggesting that on sufficiently large intervals the
principle will fail.

\medskip

We first prove a lemma that will be used in both cases.

\begin{lemma}
\label{lem:extra}
Let $u$ be a stationary point such that $x_1, x_2 \in
\omega_c$. Assume that 
\[
(x_1, x_2) \cap \omega_c = \emptyset.
\]
Then 
\begin{equation}
\label{eq:b}
\int_{x_1}^{x_2} u = \int_{x_1+1}^{x_2+1}u,
\end{equation}
and for any $m \in (0,x_2 - x_1)$,
\begin{equation}
\label{eq:c}
\int_{x_1}^{x_1+m}u < \int_{x_1+1}^{x_1 + 1 + m} u
\qquad \text{and} \qquad
\int_{x_2-m}^{x_2}u > \int_{x_2 -m +1}^{x_2+1} u.
\end{equation}
\end{lemma}

\begin{proof}
Since $(x_1, x_2) \cap \omega_c = \emptyset$,
\[
\int_{x_1+m}^{x_1+m+1} u > 0,
\]
for all $m\in(0,x_2-x_1)$, which implies
\begin{align*}
\int_{x_1+1}^{x_1+m+1} u - \int_{x_1}^{x_1+m} u 
  &= \int_{x_1}^{x_1+1} u+ \int_{x_1+1}^{x_1+m+1}u - \int_{x_1}^{x_1+m}u \\
&= \int_ {x_1+m}^{x_1+m+1}u > 0.
\end{align*}
The other two assertions are handled similarly.
\drop{
\begin{equation}
\label{ineq:extra}
0 = \int_{x_1}^{x_1 + 1}u < \int _{x_1 + m}^{x_1 + 1 + m} u,
\end{equation}
for all $0 < m < x_2 - x_1$. If $m < 1$, then
\[
0 = \int_{x_1}^{x_1 + m}u + \int_{x_1 + m}^{x_1 + 1}u
 < \int_{x_1 + m}^{x_1 + 1}u + \int_{x_1 + 1}^{x_1 + 1 + m}u.
\]
If $m > 1$, then
\[
\int_{x_1}^{x_1 + m}u = 
\int_{x_1}^{x_1 + 1}u + \int_{x_1 + 1}^{x_1 + m}u <
\int_{x_1 + 1}^{x_1 + m}u + \int_{x_1 + m}^{x_1 + m + 1}u 
= \int_{x_1 + 1}^{x_1 + 1 + m}u.
\]
In both cases (and trivially if $m = 1$),
\begin{equation}
\label{eq:a}
\int_{x_1}^{x_1 + m}u < \int_{x_1 + 1}^{x_1 + 1 + m}u
\end{equation}
for all $0 < m < x_2 - x_1$, thus proving the second assertion. 
The first assertion follows
from the same argument, replacing $m$ by $x_2-x_1$, in which case the
inequality in~\pref{ineq:extra} becomes an equality.
} % end drop
\end{proof}

\bigskip

%############ THEOREM 1 #################

\begin{theorem}
\label{thm:1}
Let $u$ be a stationary point, and assume that 
\begin{equation}
\label{def:N}
Nu:= -2a(u)u'' -a'(u){u'}^2 + b'(u)
\end{equation}
satisfies the comparison principle. Then $\omega_c$ is connected.
\end{theorem}

\begin{proof}
We proceed by contradiction.  Since $\omega_c$ is closed,
non-connectedness implies the existence of $x_1, x_2 \in \omega_c$
such that $(x_1, x_2) \cap\omega_c = \emptyset$.

Set $v = u - \tau u$. Then $v(x_1) = v(x_2) = 0$ by
Lemma~\ref{lem:ctct-periodicity}, $\int_{x_1}^{x_2}v = 0$ by
(\ref{eq:b}), and
\begin{equation}
\label{eq:int_v}
\int_{x_1}^{x_1 + m}v < 0 \text{ for all } 0 < m < x_2 - x_1
\end{equation}
by~\pref{eq:c}. Hence there exists an $\bar{x} \in (x_1, x_2)$ such that
$v(\bar{x}) = 0$.

From $(x_1, x_2) \cap \omega_c =\emptyset$ it follows that $\supp f
\cap (x_1, x_2) = \emptyset$. Hence $g = \int_{x-1}^x f$ is a
decreasing function on $(x_1, x_2)$ and $\tau g$ is an increasing
function on this interval by previous arguments. Hence $g -\tau g$ is
a decreasing function on $(x_1, x_2)$. There are three possibilities,
each leading to a contradiction with the comparison principle.

Case 1: $g \ge \tau g$ on $(x_1, x_2)$. On $(x_1,\bar{x})$,
\[
\left\{
\begin{array}{l}
Nu=g \ge \tau g = N\tau u,\\
u(x_1) = \tau u(x_1),\\
u(\bar{x}) = \tau u (\bar x).
\end{array}
\right.
\]
By the comparison principle, $u \ge \tau u$ on $(x_1,
\bar{x})$, i.e.~$v \ge 0$. But this contradicts~(\ref{eq:int_v}).

Case 2: there exists an $\tilde{x}$ such that $g \ge \tau g$ on $(x_1,
\tilde{x})$ and $g \le \tau g$ on $(\tilde{x}, x_2)$. If $\tilde{x}
\ge {x}_1$, the same argument applies. If $\tilde{x} < \bar{x}$, we
consider $(\bar{x}, x_2)$ instead, and apply the same argument. Now we
conclude $v \le 0$ on $(\bar x, x_2)$. But observe that from
$\int_{x_1}^{\bar{x}} v < 0$ by (\ref{eq:int_v}) and $\int_{x_1}^{x_2}
v = 0$ we have $\int_{\bar x}^{x_2} v> 0$, which again implies a
contradiction.

Case 3: $g \le \tau g$ on $(x_1, x_2)$. Again we obtain 
a contradiction from considering the interval $(\bar{x}, x_2)$.
\end{proof}

\bigbreak

For the second case we limit ourselves to global minimizers. The
results of this theorem do apply to the functions $a$ and $b$ given
in~\pref{def:ab}.

%############ THEOREM 2 #################

\begin{theorem}
\label{thm:2}
Let $u$ be a minimizer. Assume that $a$ and $b$ are of class $C^1$ and that
$a$ is strictly positive. Then $\omega_c$ is connected.
\end{theorem}

\begin{proof}
As in the proof of Theorem~\ref{thm:1} we assume that there exist
$x_1, x_2 \in \omega_c$ with $(x_1, x_2) \cap \omega_c = \emptyset$ to
force a contradiction. Then
\begin{equation}
\label{eq:f}
\supp f \cap (x_1, x_2) = \emptyset,
\end{equation}
and hence $g$ is a decreasing function on $(x_1, x_2)$, and an
increasing function on $(x_1 +1, x_2 + 1)$.  Now consider the
following two new functions
\[
v(x) = \left\{
\begin{array}{ll}
u(x) & \mathrm{on}\ [0,x_1],\\
u(x+1) & \mathrm{on}\ [x_1,x_2],\\
u(x) & \mathrm{on}\ [x_2,T],
\end{array}
\right.
\]
and
\[
w(x) = \left\{
\begin{array}{ll}
u(x) & \mathrm{on}\ [0,x_1+1],\\
u(x-1) & \mathrm{on}\ [x_1+1,x_2+1],\\
u(x) & \mathrm{on}\ [x_2+1,T].
\end{array}
\right.
\]
Both are admissible, i.e.\ $v,w\in K$: they are continuous by
Lemma~\ref{lem:ctct-periodicity}, implying that $v,w\in X$, and the
fact that $Bv,Bw\geq 0$ follows from Lemma~\ref{lem:extra}.  In fact
we need certain strict inequalities, which we derive after introducing
some notation.

The functions $v$ and $w$ are minimizers. To show this, write
\[
F(u|_{[x_1,x_2]}) = \int_{x_1}^{x_2} \bigl[\,a(u){u'}^2 + b(u)\,\bigr].
\]
Then since $u$ is a minimizer, and since $u$ and $v$ only
differ on $[x_1,x_2]$,
\[
F(u|_{[x_1,x_2]})\le F(v|_{[x_1,x_2]}) = F(u|_{[x_1+1,x_2+1]}),
\]
and similarly
\[
F(u|_{[x_1+1,x_2+1]})\le F(w|_{[x_1+1,x_2+1]}) = F(u|_{[x_1,x_2]}).
\]
This implies that
\[
F(u|_{[x_1,x_2]}) =  F(u|_{[x_1+1,x_2+1]}),
\]
and that $F(u) = F(v) = F(w)$. Every minimizer is also a stationary
point, and hence for $v$ and $w$ there exist positive Radon measures
$f_v$ and $f_w$ such that $\supp f_v \subset \omega_c(v)$ and $\supp
f_w \subset \omega_c(w)$.  We also denote $g_v(x) = \int_{x-1}^{x}f_v$
and $g_w(x) = \int_{x-1}^x f_w$.

For any $x \in (x_1, x_2)$, 
\begin{equation}
\label{eq:int_u}
\int_x^{x+1}u > 0.
\end{equation}
Let first $x_2 \geq x_1 + 1$. Then for any $x \in (x_1 - 1, x_1)$, 
\begin{equation*}
\begin{split}
\int_x^{x+1} v & = \int_x^{x_1} v + \int_{x_1}^{x+1} v\\
& = \int_{x}^{x_1} u + \int_{x_1+1}^{x+2} u\\
%& = \int_{x}^{x_1} u + \int_{x_1+1}^{x+2} u\\
& > \int_x^{x_1}u + \int_{x_1}^{x+1} u \qquad \text{by Lemma \ref{lem:extra}}\\
&= \int_x^{x+1} u\ge 0.
\end{split}
\end{equation*}
For any $x \in (x_1, x_2-1)$ the same is true:
\[
\int_x^{x+1} v = \int_{x+1}^{x+2} u >0,
\]
since $x+1 < x_2$, which allows us to use (\ref{eq:int_u}).
Now let $x_2 < x_1 + 1$. Then for any $x \in (x_1 -1, x_2 -1)$,
we can repeat the first argument above to conclude
\[
\int_x^{x+1} v > 0.
\]
Combining these statements we find
\[
\int_x^{x+1} v > 0 \text{ for all } x\in (x_1 -1, x_2-1),
\]
which implies $\omega_c(v) \cap (x_1 - 1, x_2 - 1) = \emptyset$. Hence
$\supp f_v \cap (x_1 - 1, x_2 - 1) = \emptyset$. But since $u$ and $v$
conincide on $[0,x_1]$, we have $g_v|_{[0,x_1]} = g_u|_{[0,x_1]}$, so
that $\supp f_u \cap (x_1 -1, x_2 - 1)=\emptyset$. Combined with
(\ref{eq:f}), this implies that $g_u|_{[x_1, x_2]}$ is constant. By
symmetry the same is true for $g_u|_{[x_1 +1, x_2 + 1]}$. Note that if
$x_2 > x_1+1$, then the overlap implies that the two constants are the
same; for the other case we now prove this.

Define $z=u-\tau u$;
the function $z$ solves the equation
\begin{multline}
-2a(u)z'' = g_u-\tau g_u + \{a'(u){u'}^2 - a'(\tau u){(\tau u)'}^2 \} \\
  - \{ b'(u) - b'(\tau u)\} 
  + \{2a(u)-a(\tau u)\}(\tau u)''
\label{eq:z}
\end{multline}
on the interval $(x_1,x_2)$. Of the right-hand side, we have seen
above that the term $g_u-\tau g_u$ is constant on $(x_1,x_2)$; let us
suppose it non-zero for the purpose of contradiction.  The function
$z$ is of class $C^1$, and both $z$ and $z'$ vanish at $x=x_{1,2}$.
Therefore the assumed regularity on $a$ and $b$ implies that the
expressions between braces are continuous on $[x_1,x_2]$ and zero at
$x=x_{1,2}$. The sign of the right-hand side of~\pref{eq:z} is
therefore determined by $g_u-\tau g_u$, and most importantly, is the
same at both ends $x_1$ and $x_2$; therefore the sign of $z$, at
$x=x_1+$ and $x=x_2-$, is also the same. This contradicts the
following consequence of Lemma~\ref{lem:extra}:
\[
\int_{x_1}^{x_1+m} z < 0  \qquad \text{and}\qquad
\int_{x_2-m}^{x_2} z > 0 \qquad \text{for all}\quad 0<m<x_2-x_1.
\]

This leaves $g_u = \tau g_{u}$ on $(x_1, x_2)$. But then, by
uniqueness of the initial-value problem, $u(x) = u(x+1)$ for all $x
\in [x_1, x_2]$, and $[x_1, x_2] \subset \omega_c$, contrary to our
assumption that $(x_1, x_2) \cap \omega_c = \emptyset$.
\end{proof}

\section{Symmetry}
\label{sec:symmetry}

In the introduction we raised the question whether the stationary
points or minimizers inherit the symmetry of the formulation, or to
put it differently, whether non-symmetric solutions exist.

For the discussion of this question it is useful to introduce an
equivalent formulation of the Euler-Lagrange equation~\pref{eq:EL}
similar to the Hamiltonian-systems formulation used in the proof of
Theorem~\ref{thm:minimizing_rods_intersect}. For the length of this
section we assume that Theorem~\ref{thm:structure_g_3} applies and
therefore that there is a single contact interval $[x_0,x_1]$.

\medskip

By multiplying~\pref{eq:EL} with $u'$ and integrating one finds
that the function $H$, defined by
\begin{equation}
\label{def:H}
H := -a(u){u'}^2 + b(u) - gu,
\end{equation}
is piecewise constant, and that $H$ and $g$ jump at the same values of
$x$. The function~$g$ takes three values on $[0,T]$, these being the
values $g_1$ and $g_2$ introduced in Theorem~\ref{thm:structure_g_3},
and the value $g_0=0$ outside of the extended contact interval
$[x_0,x_1+1]$.  (Note that $g_1$ and $g_2$ may be equal).  We claim
that $H$ also takes three values, $H_0$, $H_1$, and $H_2$, and that
these values correspond to those of $g$, i.e.\ that the \emph{pair}
$(g,H)$ takes three values $(0,H_0)$, $(g_1,H_1)$, and $(g_2,H_2)$
(although it may happen that $(g_1,H_1)=(g_2,H_2)$).

To prove this claim, first consider the case of $p>0$, where $p$ 
is defined as in~\pref{def:p}. Then 
\[
u|_{(x_0,x_0+p)} \equiv u|_{(x_0+1,x_0+1+p)} \qquad\text{and}\qquad
g|_{(x_0,x_0+p)} \equiv g|_{(x_0+1,x_0+1+p)}
\]
by Lemma~\ref{lem:ctct-periodicity} and~\pref{def:g1g2}.  Therefore
$H$ is the same on these two intervals. Repeating this argument for
all subintervals of $[x_0,x_1+1]$ of the form $(x_0+k,x_0+k+p)$ and
$(x_0+k+p,x_0+k+1)$ we find that $H$ takes two values on the interval
$[x_0,x_1+1]$, $H_1$ and $H_2$, and that these coincide with the
values $g_1$ and $g_2$ of $g$.

When $p=0$, a similar argument yields that $H$ takes only one value on
$[x_0,x_1+1]$ (as does $g$).

\medskip
A consequence of this characterization of $H$ is the following lemma:
\begin{lemma}
\label{lem:equal-u}
Under the conditions and notation of Theorem~\ref{thm:structure_g_3}, 
\[
u(x_0) = u(x_0+p) = u(x_0+1) = u(x_0+1+p) = \dots = 
u(x_1+1).
\]
\end{lemma}

\begin{proof}
When $p=0$ the statement follows from
Lemma~\ref{lem:ctct-periodicity}. For $p>0$, note that at any of the
\emph{interior} jump points, i.e.\ at all jump points except $x_0$ and
$x_1+1$, we have $[H] = - [g] u$ where $[H] = \pm(H_2-H_1)$ and $[g] =
\pm(g_2-g_1)$.  Regardless of the sign this equation has only one
solution $u$.  For the remaining two points $x_0$ and $x_1+1$ the
result follows from Lemma~\ref{lem:ctct-periodicity}.
\end{proof}

We still need to show that the value of $H$ is the same on both sides
of the extended contact interval $[x_0,x_1+1]$, so that we can define
the value $H_0$ unambiguously. If one of the ends of this interval
equals $0$ or $T$ there is nothing to prove; we therefore assume that
$\min\{x_0,T-x_1-1\}\geq d>0$.  Now multiply~\pref{eq:EL} with the
function
\[
v(x) = \begin{cases}
  \frac xd u'(x) & 0<x<d\\
  u'(x) & d \leq x \leq T-d \\
   \frac {T-x}d u'(x) & T-d < x< T,
\end{cases}  
\]
and integrate to find
\[
- \frac 1d \int_0^d H + \frac 1d \int_{T-d}^T H = 0.
\]
Since $H$ is constant on $(0,d)$ and on $(T-d,T)$ the two constant
values are equal; we then define $H_0$ to be this value.

\medskip

We now turn to the implications of this characterization of solutions
$(u,g)$ and the associated pseudo-Hamiltonian function $H$.

\begin{theorem}
\label{thm:symmetry}
Let $u$ be a stationary point with a single contact
interval $[x_0,x_1]$. Let~$p$ be given as in~\pref{def:p}, and 
define the set of jump points $J = \{x_0,x_0+p,x_0+1,x_0+1+p,\ldots,x_1+1\}$.
\begin{enumerate}
\item There exists $\alpha\in\R$ such that at any $x\in J$, $u'(x) = \pm\alpha$.
\end{enumerate}
Now assume that $b$ is non-decreasing on $[1,\infty)$.
\begin{enumerate}
\addtocounter{enumi}1
\item\label{thm:symmetry:b} If the operator $N$ given in~\pref{def:N} satisfies the comparison
principle, then $u$ is symmetric on $[0,T]$.
\item If $u$ is a minimizer with $u'(x_0) = -u'(x_1+1)$, 
then $u$ is symmetric on $[0,T]$.
\end{enumerate}
\end{theorem}

\begin{proof}
For the first part write 
\[
u'^2 = \frac{b(u) - gu - H}{a(u)},
\]
and note that by the proof of Lemma~\ref{lem:equal-u} the sum $gu+H$
is continuous.

\medskip

For the second part, note that by Lemma~\ref{lem:equal-u} $u$ has the
same value on each end of the interval $[x_0,x_0+p]$ (if $p>0$) or
$[x_0,x_0+1]$ (if $p=0$). By the uniqueness that follows from the
comparison principle the function $u$ is symmetric on this
interval. By repeating this argument over all subintervals of
$[x_0,x_1+1]$ we find that $u$ is symmetric on $[x_0,x_1+1]$.

The functions $u_1(t) := u(x_0-t)$ and $u_2(t) := u(x_1+1+t)$,
therefore, have the same zeroth and first derivatives at $t=0$; they
satisfy the same equation~\pref{def:H} (note that $H$ is symmetric on
$[x_0,x_1+1]$); therefore the two functions are equal as long as they
both exist. This implies that lack of symmetry must stem from a
difference in domain of definition of $u_1$ and $u_2$ for $t>0$.

We claim that neither $u_1$ nor $u_2$ has an interior
maximum. Assuming this claim, the assertion of the theorem follows
since the monotonicity of $u_{1,2}$ then implies that the boundary
condition $u_{1,2}(t) = 1$ has at most one solution $t$.

Now assume that $u_1$ has a maximum at $t_1>0$. The function $u_1$ is
solution of the Hamiltonian system~\pref{def:H}, where $H$ and $g$ are
constant for $t>0$.  As in the proof of
Theorem~\ref{thm:minimizing_rods_intersect}, therefore
$u_1(t_1)>1$. Choose a bounded interval $I\subset[0,\infty)$ such that
$u>1$ on $\Int I$ and $u(\partial I) = 1$.

The reduced functional $\tilde F(v) = \int_I [a(v){v'}^2 + b(v)]$ has
a global minimizer $\tilde v$ in the class of functions $v$ satisfying
$v(\partial I) = 1$.  From studying the perturbation $v \mapsto
\min\{v,1\}$ and using the monotonicity of $b$ it follows that $\tilde
v\leq 1$ on $I$. By the comparison principle this is the only
stationary point of $\tilde F$, a conclusion that contradicts the fact
that $u_1$ is a different stationary point.

\medskip

For the third part, first note that the support of the continuous
function $x\mapsto\int_x^{x+1} u$ is $[x_0,x_1]$; therefore
\begin{quote}
for every $\epsilon>0$ there exists $\delta>0$ such that any
perturbation $v$ with $d(\supp v,[x_0,x_1+1])>\epsilon$ is admissible
provided $\Modd{v}_{L^\infty}< \delta$.
\end{quote}
We will use this below.

The assumption on the derivatives places us in the same position as
above: the functions $u_1(t) := u(x_0-t)$ and $u_2(t) := u(x_1+1+t)$
are equal as long as they both exist. Again we will show that neither
may have an interior maximum, but by a different argument.

Assume that $u_1$ has a maximum.  By defining $t_1 = x_0$ the boundary
condition on $u$ takes the form $u_1(t_1) = 1$.  Pick
\[
\max\{1,u_1(0)\} < \beta < \max\{u_1(t): 0\leq t \leq t_1\}
\]
and define the set $S = \{t\in[0,t_1]: u_1(t) \geq \beta \}$; 
we can assume that for $\epsilon = \inf S>0$ we have
$\max\{u_1(t): 0\leq t \leq t_1\} - \beta < \delta$ for 
the associated $\delta$ given above.

Now define $v(t) = \min\{\beta,u_1(t)\}$. The function $v$ is
admissible by construction; it differs from $u_1$ only on the set $S$,
and therefore the difference in energy is given by (with a slight
abuse of notation)
\[
F(v)-F(u_1) = \int_S \bigl[-a(u_1){u_1'}^2 + b(\beta)-b(u_1)\bigr] < 0.
\]
This contradicts the assumption of minimality. 
\end{proof}

The conditions of Theorem~\ref{thm:symmetry} are quite sharp. We
demonstrate this with two examples.

\medskip

\textbf{Example 1: \boldmath $b$ is decreasing on $[1,\infty)$.}
It is relatively straightforward to construct a non-symmetric
stationary point by choosing an appropriate function $b$ that is
decreasing on $[1,\infty)$, thus showing that
part~\ref{thm:symmetry:b} of Theorem~\ref{thm:symmetry} is sharp.

\begin{figure}[ht]
\centering
\centerline{\psfig{figure=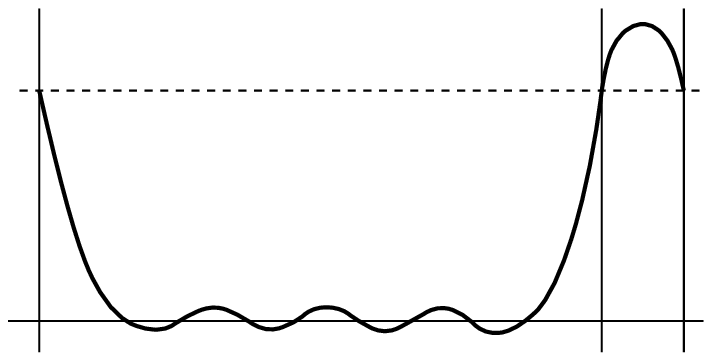,height=3.5cm}}
\vskip3mm
\caption{A non-symmetric stationary point can be constructed by
defining $b(u)$ appropriately for $u>1$.}
\label{fig:nonsym1}
\end{figure}

Take a symmetric stationary point $u$ for which $u\leq 1$ on $[0,T]$,
$u'(T)>0$, and for which the contact set is bounded away from $x=T$
(see the next section for examples). Close to $x=T$, the function $u$
satisfies
\[
{u'}^2 =  \frac{b(u)-H}{a(u)}
\]
for some $H\in\R$, and since $u'(T)>0$, $b(1)>H$. Now change $b(u)$
for $u>1$ such as to have (for instance) $b(2)=H$, and continue the
solution $u$ past $x=T$. By construction $u(T+\tilde T)=2$, for some
$\tilde T>0$, and $u'(T+\tilde T)=0$; by symmetry then $u(T+2\tilde T)
= 1$. The new function $u$ defined on the domain $[0,t+2\tilde T]$ is
a non-symmetric stationary point (Figure~\ref{fig:nonsym1}).

\medbreak
\textbf{Example 2: \boldmath equal (non-opposite) derivatives 
on $\partial \omega_c$.}  For certain functions $b$ and domains
$[0,T]$ global minimization favours breaking of symmetry. We
demonstrate this for the functional
\[
F(u) = \int\bigl[\, {u'}^2 + \alpha (1-u^2)^2\, \bigr],
\]
where $\alpha$ will be chosen appropriately. We consider the
functional $F$ on functions $u:[0,1]\to\R$ with boundary conditions
$u(0)=u(1)=0$; although this is slightly different from the setup in
the rest of the paper, it simplifies the argument, and the extension
to a more general situation is intuitively clear.

We will show that
\begin{equation}
\label{ineq:asymmetry}
\inf \left\{ F(u): \int u \geq 0\right\}  
\;<\; \inf \left\{F(u): \int u \geq 0 \text{ and } 
u \text{ is symmetric}\right\}.
\end{equation}
The infimum on the right-hand side is bounded from
below,
\[
F(u) \geq (1-\alpha c/2)\int {u'}^2 + \alpha,
\]
by the Poincar\'e inequality
\begin{equation}\label{ineq:Poincare}
\int_0^d u^2 \leq cd^2 \int_0^d {u'}^2 
\qquad \text{for all $u$ with $u(0)=0$ and $\int u = 0$}.
\end{equation}
The function $v(x) = a+\cos(b(1-x/d))$ is optimal in this inequality,
where $a\simeq 0.22$ and $b\simeq 4.49$ are determined by the boundary
condition $v(0)=0$ and the integral condition $\int v = 0$. The
Poincar\'e constant equals $c\simeq0.0495$.  Note that for symmetric
functions $u$ we may take $d=1/2$.

At the function $w(x) = \sin2\pi x$ the functional $F$ has
the value $F(w) = 2\pi^2 + 3\alpha/8$.
For all $\alpha\in(16\pi^2/5,2/c] \simeq (31.6,40.3]$ therefore
\[
F(w) = 2\pi^2 + 3\alpha/8 < \alpha \leq \inf \left\{F(u): \int u \geq
0 \text{ and } u \text{ is symmetric}\right\},
\]
which demonstrates~\pref{ineq:asymmetry}.

\begin{figure}[ht]
\centering
\centerline{\psfig{figure=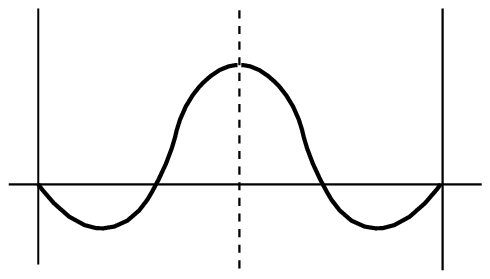,width=4cm}\qquad
  \psfig{figure=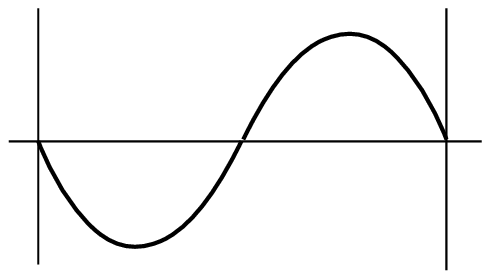,width=4cm}}
\caption{Under symmetry conditions the effective domain,
the domain on which $\int u = 0$, is half the actual
domain size. Equivalently, more (costly) oscillations are
necessary.}
\label{fig:asymmetry}
\end{figure}

The reason for this preference for asymmetry can be recognized in the
constant in the Poincar\'e inequality~\pref{ineq:Poincare} (see
Figure~\ref{fig:asymmetry}). For symmetric functions the relevant
class is $\{ u:[0,1/2]\to\R: u(0) = \int u =0\}$, and for more general
functions $\{ u:[0,1]\to\R: u(0) = u(1) = \int u = 0\}$.  For this
latter class the Poincar\'e coefficient is achieved by the function
$w$ above with the value $\overline c = 1/4\pi^2\simeq 0.0253$, which
is larger than $c(1/2)^2 = 0.0124$.

\section{Numerical simulations}
\label{sec:numerics}

In this section we describe in detail our numerical simulations
of stationary points of $F$ under constraint, i.e.\ of solutions of
\begin{align}
\label{num:eq:ode1}
&-2a(u)u''-a'(u){u'}^2 + b'(u) = \int_{x-1}^x f,\\
\label{num:eq:bc}
&u(0) = u(T) = 1,\\      
&\supp f \subset \omega_c,\\
&f\ \mathrm{a\ positive\ Radon\ measure},\\
&\int_x^{x+1} u\ge 0\quad\forall x\in[0,T-1].
\label{num:ineq}
\end{align}
We concentrate on the case in which the solution is symmetric and the
contact set is non-empty, and we use the fact that the right-hand side
in the differential equation can be characterised explicitly
(see~\pref{eq:g_explicit}).  We further simplify by replacing the
inequality~\pref{num:ineq} by the condition that the function
$x\mapsto \int_x^{x+1} u$ has a second-degree zero at $x=x_0$, leading
to the new system in the unknowns $(u,x_0,G)$
\begin{align}
\label{num:eq:ode2}
&-2a(u)u''-a'(u){u'}^2 + b'(u) =  g(x;x_0,T-x_0-1,G),\\
\label{num:eq:bc2}
&u(0) = u(T) = 1,\\      
&u(x_0)=u(x_0+1),\\
&\int_{x_0}^{x_0+1} u = 0.
\label{num:int}
\end{align}
For brevity we shall write $\g(x;x_0,T,G)$ for $g(x;x_0,T-x_0-1,G)$.

\begin{lemma}
Assume that the operator on the left-hand side of~\pref{num:eq:ode2}
satisfies the comparison principle.  Then any solution of
problem~(\ref{num:eq:ode1}-\ref{num:ineq}) with non-empty contact set
is also a solution of~(\ref{num:eq:ode2}-\ref{num:int}); vice versa,
any solution of~(\ref{num:eq:ode2}-\ref{num:int}) is also a solution
of~(\ref{num:eq:ode1}-\ref{num:ineq}).
\end{lemma}

\begin{proof}
Since the implication
$\text{(\ref{num:eq:ode1}-\ref{num:ineq})}\Longrightarrow
\text{(\ref{num:eq:ode2}-\ref{num:int})}$ follows by construction, it
suffices to show the opposite implication; in fact, since an
admissible measure $f$ can be constructed from any $\g(x;x_0,T,G)$, it
is sufficient to show that solutions
of~(\ref{num:eq:ode2}-\ref{num:int}) satisfy
\[
\int_x^{x+1} u \ge 0\qquad \forall x\in[0,T-1].
\]
We show slightly more, namely that
\[\int_x^{x+1} u = 0\qquad \forall x \in [x_0,T - x_0 - 1]
\]
and that 
\[
\int_x^{x+1} u > 0\qquad \forall x \notin [x_0,T - x_0 - 1].
\]

The function $u$ is symmetric by Theorem~\ref{thm:symmetry}. Since
$u(x_0) = u(x_0 + 1)$,
\[
u(x_0) = 
u(x_0 + 1) = u(T - x_0) = u(T - x_0 - 1) =: \bar u.
\]
Set $u_1(x) = u(x_0+x)$, and $u_2(x) = u(x_0 +x + 1)$ for all $x \in
[0,T - 2x_0 - 1]$. By construction, $\g(x;x_0,T,G) = \g(x+1;x_0,T,G)$
for all $x \in [x_0, T - x_0 -1]$. Hence, if we set $h(x) = \g(x +
x_0;x_0,T,G)$, for all $x \in [0, T - 2x_0 - 1]$, then $u_1$ and $u_2$
both satisfy
\[
\begin{array}{l}
-2a(v)v'' - a'(v){v'}^2 + b'(v) = h,\\
v(0) = v(T-2x_0 - 1) = \bar u,
\end{array}
\]
By uniqueness, $u_1 = u_2$ on $[0,T - 2x_0 - 1]$. In terms of $u$ this
means $u(x) = u(x+1)$ for all $x \in [x_0, T - x_0 - 1]$.
But that implies that 
\[
\int_x^{x+1}u = 0\qquad \forall x\in [x_0,T - x_0 - 1].
\]

It remains to be shown that 
\begin{equation}
\label{ineq:lem}
\int_x^{x+1}u > 0\qquad \forall x\notin [x_0,T - x_0 - 1].
\end{equation}
By symmetry we only show this for $x<x_0$. Let $u_p$ and $g_p$ be the
$1$-periodic extrapolation of $u|_{[x_0,x_0+1]}$ and
$g|_{[x_0,x_0+1]}$; note that $\int_x^{x+1} u_p = 0$ for every
$x$. For $x<x_0$,
\begin{equation}
\label{ineq:lem2}
-2a(u)u''-a'(u){u'}^2 + b'(u) = 0 < g_p =
-2a(u_p)u_p''-a'(u_p){u_p'}^2 + b'(u_p),
\end{equation}
implying that $u>u_p$ for $x=x_0-$ and therefore also~\pref{ineq:lem}
for $x=x_0-$. If $u$ and $u_p$ intersect again at some $\tilde x<x_0$,
then the comparison principle and~\pref{ineq:lem2} imply that $u\leq
u_p$ on $[\tilde x,x_0]$, in contradiction with the previous
statement.  This concludes the proof.
\end{proof}

We discuss two different ways of calculating solutions of the 
problem~(\ref{num:eq:ode2}-\ref{num:int}).

\subsection{Continuation}

We implemented a strategy of continuation of solutions, using the
continuation package \AUTO~\cite{AUTO}, and we chose the simple case
\begin{equation}
\label{def:simple_case}
a(u) = \frac12, \qquad b(u) = \frac12 (u+1)^2.
\end{equation}
To implement system~(\ref{num:eq:ode2}-\ref{num:int}) in \AUTO, we
divide $[0,T]$ into three subdomains, $[0,x_0]$, $[x_0, x_0+1]$ and
$[x_0+1,T]$ and specify the equations
\begin{align}
\label{eq:AUTO1}
\left.
\begin{array}{r}
-u_1''(x_1) + u_1(x_1) + 1 = \g(x_1;x_0,T,G)\\
x_1' = 1
\end{array}
\right\}
&\ \mathrm{on\ } [0,x_0],\\
\label{eq:AUTO2}
\left.
\begin{array}{r}
-u_2''(x_2) + u_2(x_2) + 1 = \g(x_2;x_0,T,G)\\
x_2' = 1
\end{array}
\right\}
&\ \mathrm{on\ } [x_0, x_0+1],\\
\label{eq:AUTO3}
\left.
\begin{array}{r}
-u_3''(x_3) + u_3(x_3) + 1 = \g(x_3;x_0,T,G),\\
x_3' = 1
\end{array}
\right\}
&\ \mathrm{on\ } [x_0+1,T],
\end{align}
with boundary conditions
\begin{gather}
u_1(0) = 1,\nonumber\\
u_1(x_0) = u_2(x_0),\ u_1'(x_0) = u_2'(x_0),\nonumber\\
u_2(x_0) = u_3(x_0),\
u_2'(x_0) = u_3'(x_0), \label{eq:BCs}\\
u_3(T) = 1,\nonumber\\
u_2(x_0) = u_2(x_0+1),\nonumber\\
x_1(0)=0,\ x_2(x_0)=x_0,\ x_3(x_0+1)=x_0+1.\nonumber
\end{gather}
and integral condition
\begin{equation}
\label{eq:ICs}
\int_{x_0}^{x_0+1} u_2 = 0.
\end{equation}
Note that in (\ref{eq:AUTO1})--(\ref{eq:AUTO3}) we have added trivial
equations in order to solve for the $x_i$ variables, which are
required in the evaluation of $\bar{g}(x;x_0,G,T)$.
 
There are still some technicalities that have to be overcome: \AUTO\
is not well-equipped to handle systems with a discontinuous right-hand
side, such as the function $g(x;x_0,G,T)$ that is supplied here. We
remedy this by using a low-order method for all simulations, and we
smooth the function $g$ given in (\ref{eq:g_explicit}) by substituting
$\arctan$s for Heaviside functions:
\begin{multline*}
\tilde{g}(x;x_0,T,G) = \\
\left\{
\begin{array}{l}
\frac{g_1}{\pi}\big(\arctan(A(x - x_0)) - 
\arctan(A(x - T - x_0))\big) 
\quad \quad \hfill \mathrm{if\ } T - 2x_0 \notin \N,\\  
\quad +\frac{(g_2 - g_1)}{\pi}\sum_{i=1}^P \big[\arctan(A(x - X_i)) 
-\arctan(A(x - Y_i))\big],\\
\frac {1}{\pi}\big(\arctan(A(x - x_0))- \arctan(A(x - T  -x_0))\big)
\quad \quad \hfill \mathrm{if\ } T - 2x_0 \in \N,
\end{array} 
\right.
\end{multline*}
where $X_i$ and $Y_i$ are as in (\ref{eq:XiYi}).
In the limit $A \to \infty$, $\tilde{g}(x;x_0,T,G)$ converges
pointwise to $\g(x;x_0,T,G)$.

There are nine differential equations with ten boundary conditions and
one integral condition. This means that we expect to specify three
free parameters to obtain a one-parameter curve of solutions. These
are $T$, $x_0$, and an additional parameter $\beta$.  It worked well
to choose the freedom in $\beta$ in modulating the values of
$g_{1,2}$:
\[
\tilde{g}_1 = g_1 + \beta\qquad \text{and}\qquad 
\tilde{g}_2 = g_2 + \beta.
\]
One may prove \emph{a priori} that $\beta = 0$ by remarking that 
\[
\int_{x_0}^{x_0+1}\tilde{g} = \int_{x_0}^{x_0+1}\bigl[-u'' + u + 1\bigr] = 1,
\]
and using (\ref{eq:g1g2}) to find
\[
1 = \int_{x_0}^{x_0+1} \tilde{g} = p(g_1 + \beta) + (1-p)(g_2 + \beta)
= 1 + \beta.
\]
We have found no other role for $\beta$ than
to accommodate for small numerical inaccuracies due to the
discontinuous right-hand side. In all simulations $\beta \simeq 10^{-4}$.

\begin{figure}[hbtp]
\pspicture(12,9.4)
\rput(6,4.7){
\includegraphics[angle=-90,width=11cm]{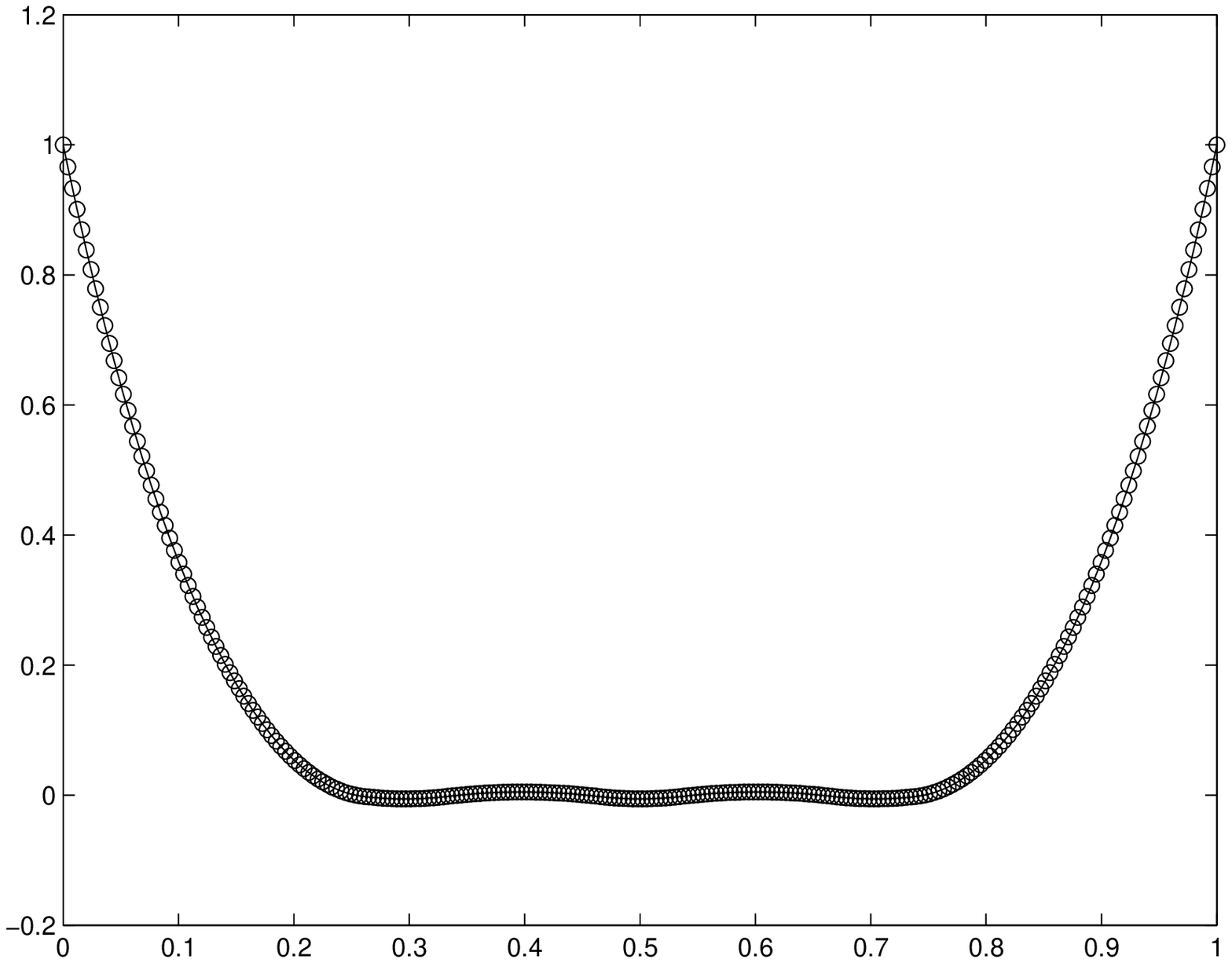}
}
\uput[l](.7,6){$u$}
\uput[d](7.7,0.3){$x$}
\endpspicture
\caption{A comparison of a solution of system
(\ref{eq:AUTO1})--(\ref{eq:ICs}) produced with \AUTO\ ($\circ$
symbols) to an explicit solution, for a generic value of $T$ (here
scaled to 1): $T =4.91635$. In this simulation $A = 1000$.}
\label{fig:compare}
\end{figure}

We have validated the code by comparing solutions from \AUTO\ with 
explicit solutions. An example is given in Figure \ref{fig:compare}.

\begin{figure}[!hbtp]
\pspicture(0,0)(12,9.4)
\rput(6,4.7){
\includegraphics[width=11cm]{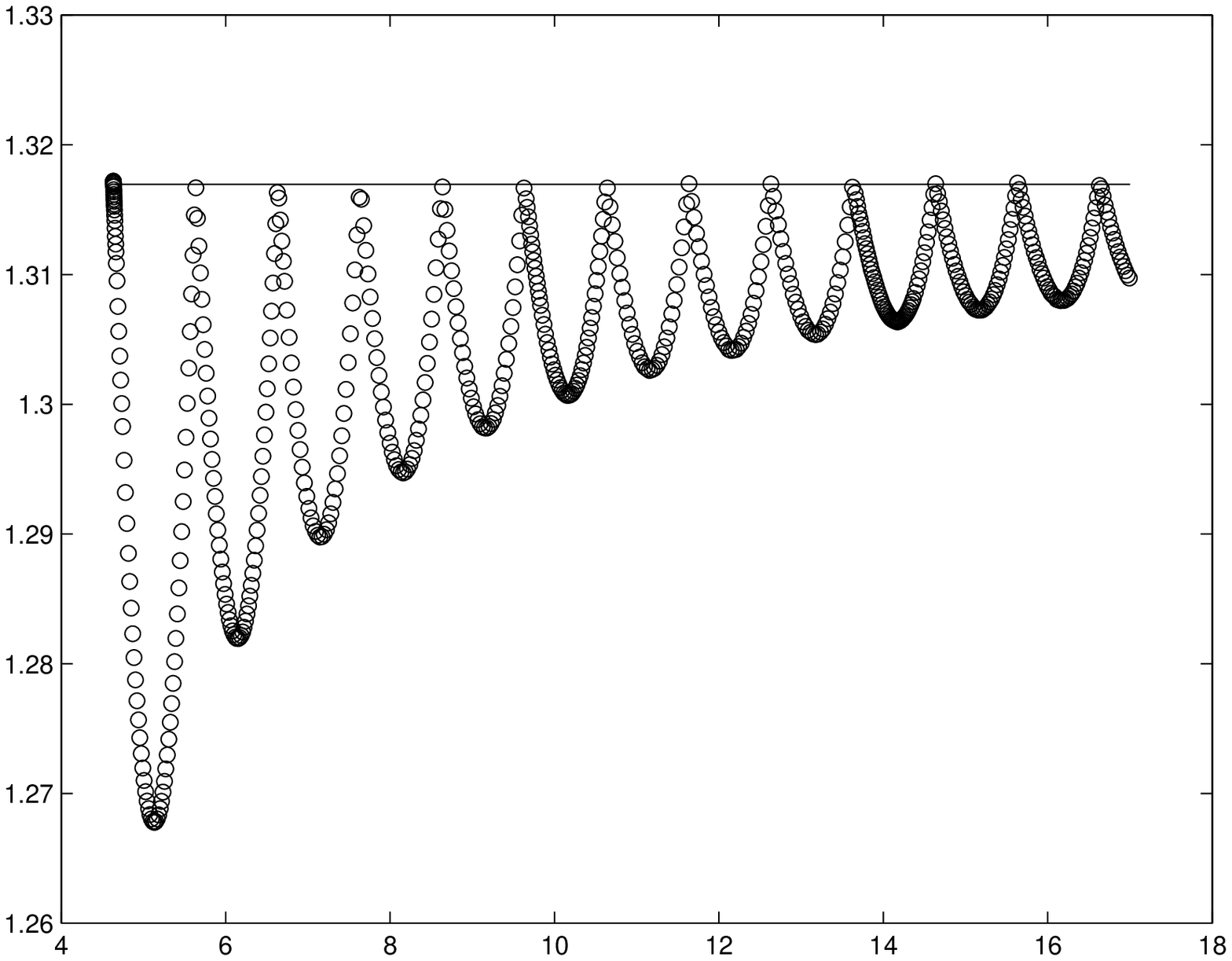}
}
\uput[l](.7,5.4){$x_0$}
\uput[d](7.5,0.3){$T$}
\endpspicture
\caption{Behaviour of $x_0$ as a function of domain size $T$ for system
(\ref{eq:AUTO1})--(\ref{eq:ICs}) computed with \AUTO. As $T$ grows,
$x_0$ remains bounded and converges to $\log(2 + \sqrt 3)$ (horizontal
line). } 
\label{fig:x0t}
\end{figure}

As we have seen in the discussion at the beginning of this Section, as
$T$ becomes larger the minimizer $u$ has to have a contact point, and
for large enough values even a full interval of contact. The point
$x_0$, the leftmost point of contact, is determined as part of the
solution; one may wonder how this point depends on $T$. For operators
$N$ that satisfy the comparison principle, it is straightforward to
prove that $x_0$ remains bounded for all $T$. Moreover, for the
operator under consideration here, as $T \to \infty$, $x_0 \to \log(2
+ \sqrt 3)$. These two phenomena are illustrated in
Figure~\ref{fig:x0t}.

\begin{figure}[!hbtp]
\pspicture(12,9.4)
\rput(6,4.7){
\includegraphics[width=11cm]{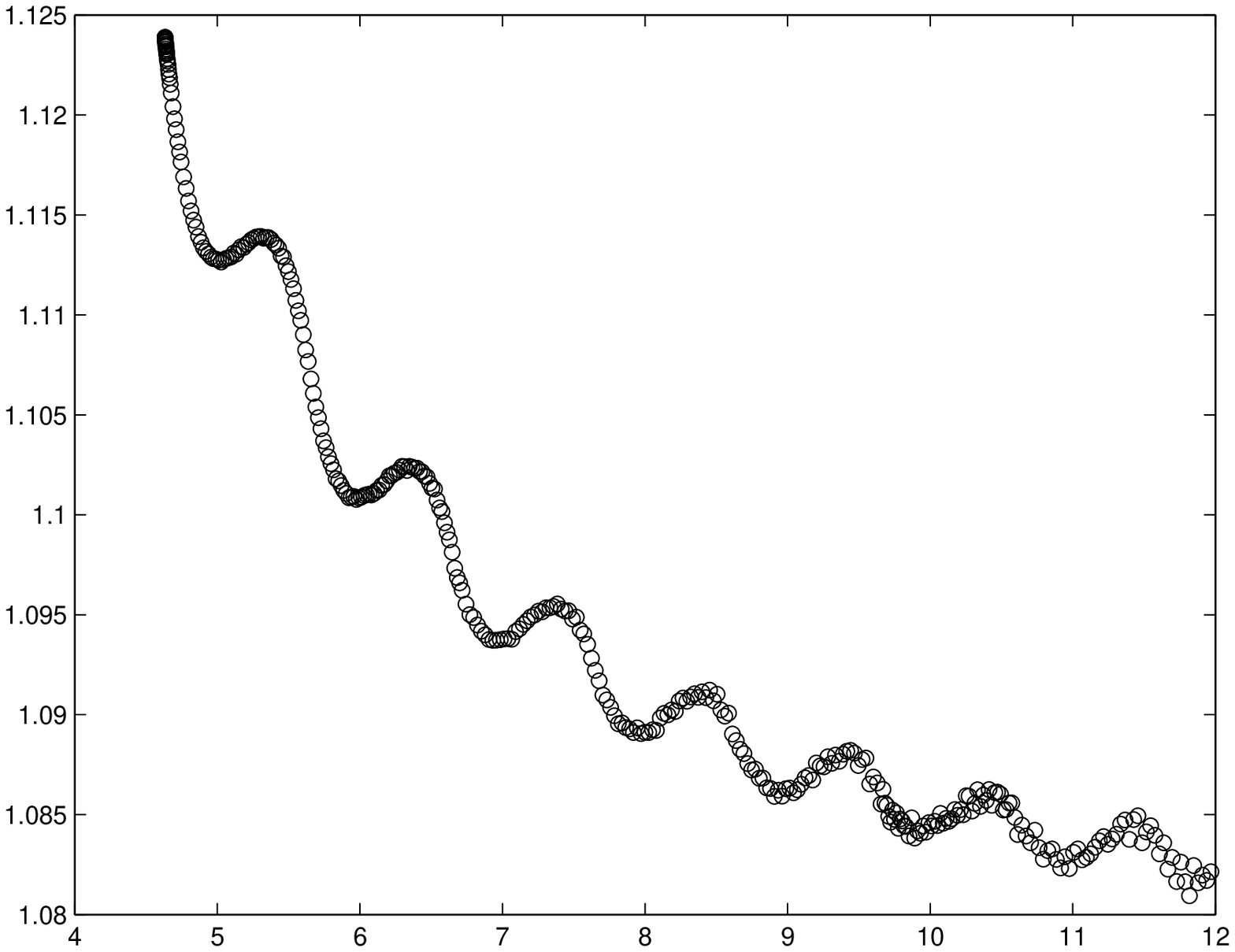}
}
\uput[l]{90}(0.5,6){$F(u)-T/2$}
\uput[d](8.2,0.5){$T$}
\endpspicture
\caption{Behaviour of $F(u) - T/2$ as a function of domain size $T$
for system (\ref{eq:AUTO1})--(\ref{eq:ICs}) computed with \AUTO. As $T$
grows, $F(u) - T/2$ oscillates towards 1, the energy of $u \equiv 0$
on a unit length interval.}
\label{fig:FvsT}
\end{figure}

Since $|g_1-g_2|\to0$ as $P$ (and therefore $T$) increases, $g$ becomes
constant in the limit of large $T$. By the comparison principle,
$u$ does the same, implying that $F(u)/T \to 1$. The start of the
convergence to 1 is shown in Figure \ref{fig:FvsT}.

\subsection{Directly solving the boundary-value problem}
\begin{figure}[!hbtp]
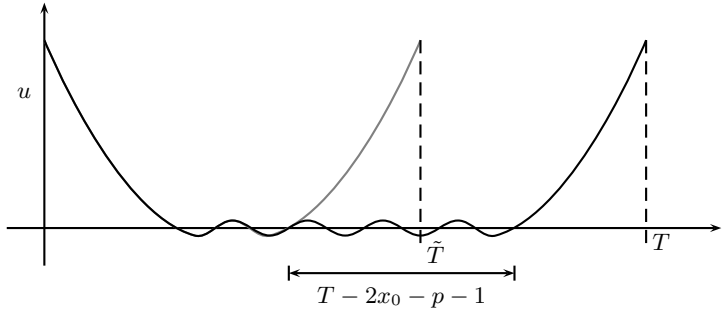

\pspicture(-1,-1)(11,3)
%\psgrid(0,0)(-1,-1)(9,3)

\rput(1,0){
\psline{->}(0,-.5)(0,3)
\psline{->}(-.5,0)(9,0)

\pscurve[linecolor=gray](0,2.5)(2,-.1)(2.5,.1)
(3,-.1)(5,2.5)

\pscurve(0,2.5)(2,-.1)(2.5,.1)
(3,-.1)(3.5,.1)
(4,-.1)(4.5,.1)
(5,-.1)(5.5,.1)
(6,-.1)(8,2.5)

\psline[linestyle=dashed](5,2.5)(5,-.2)
\uput[d](5.2,0){$\tilde T$}

\psline[linestyle=dashed](8,2.5)(8,-.2)
\uput[d](8.2,.1){$T$}
\uput[l](0,1.8){$u$}

\psline{<->}(3.25,-0.6)(6.25,-.6) 
\psline(3.25,-.5)(3.25,-.7)
\psline(6.25,-.5)(6.25,-.7)
\uput[d](4.75,-.6){$T - 2x_0 -p -1$}
}

\endpspicture
\caption{Schematic picture of the idea behind $\tilde T = 2x_0 + p +
1$. Since $u$ (solid black line) is periodic between $x_0$ and $T -
x_0$, we can cut out an interval of length $T - 2x_0 - p -1$ and find
the corresponding solution on $[0,\tilde T]$. }
\label{fig:reduction}
\end{figure}

Computing solutions of the rod equations---rather than the 
simpler problem~\pref{def:simple_case}---using \AUTO\ has proved
difficult, for reasons that we do not understand well. 
Instead, a boundary-value problem solver from Matlab was
used to create Figure \ref{fig:typical}. Set
\[
Lu = -\frac{2u''}{4\pi^2(1+u^2)^{\frac 52}} +
\frac{5u{u'}^2}{4\pi^2(1+u^2)^ 
\frac 72} - \frac{3u}{r^2(1 + u^2)^{\frac 52}} +  
\frac{\alpha}{(1+u^2)^{\frac 32}}.
\]
To find a solution of 
\[
Lu = g(x;x_0,T,G),\quad u(x_0) = u(x_0 + 1),\quad 
\int _{x_0}^{x_0+1}u = 0,
\]
for a generic value of $T$ (large enough) we construct a two-parameter
shooting problem. Fix $G$ and $x_0$ and consider the boundary-value problem
\begin{equation}
\left.
\begin{array}{ll}
\label{eq:MATLAB1}
Lu_1 = 0   &\ \mathrm{on\ } [0,x_0],\\
Lu_2 = g_1 &\ \mathrm{on\ } [x_0, x_0+p],\\
Lu_3 = g_2 &\ \mathrm{on\ } [x_0+p,x_0+1],\\
Lu_u = g_1 &\ \mathrm{on\ } [x_0 + 1, x_0 + 1 + p],\\
Lu_5 = 0   &\ \mathrm{on\ } [x_0+ 1 + p,\tilde T],
\end{array}
\right\}
\end{equation}
with boundary conditions
\begin{gather}
u_1(0) = 1,\\
u_1(x_0) = u_2(x_0),\ u_1'(x_0) = u_2'(x_0),\\
u_2(x_0+p) = u_3(x_0+p),\ u_2'(x_0+p) = u_3'(x_0+p),\\
u_3(x_0+1) = u_4(x_0+1),\ u_3'(x_0+1) = u_4'(x_0+1),\\
u_4(x_0+1+p) = u_5(x_0+1+p),\ u_4'(x_0+1+p) = u_5'(x_0+1+p),\\
u_5(\tilde{T}) = 1.\label{eq:MATLAB10}
\end{gather}
Here, as before, $p \equiv T - 2x_0 -1 \text{ (mod 1)}$, 
$P = \min\{n\in \N:\ n\ge T - 2x_0 - 1\}$, and 
\[
g_1 = \frac{GP}{P+1-p},\quad g_2 = \frac{G(P+1)}{P+1-p},
\]
by Theorem \ref{thm:structure_g_3}.  Note that this is not exactly the
same problem as~(\ref{num:eq:ode2}-\ref{num:eq:bc2}), since the
periodic section has been reduced from $P$ periods to a single period,
and the solution is defined correspondingly on a smaller domain of
length
\[
\tilde T = 2x_0 + p + 1.
\]
This allows us to use the decomposition in five subdomains for any
$T$, which facilitates computation. This is illustrated in Figure
\ref{fig:reduction}.

We now vary $x_0$ and $G$ to find solutions of system
(\ref{eq:MATLAB1})--(\ref{eq:MATLAB10}) that satisfy
\[
u_2(x_0) = u_3(x_0+1),\quad \int_{x_0}^{x_0+p} u_2 + 
\int_{x_0 + p}^{x_0 + 1}u_3 = 0,
\]
using a standard Matlab boundary-value problem solver, \texttt{bvp4c}.
An example solution is drawn in Figure~\ref{fig:typical} in which we
have used $\alpha = 1/2\pi,\ r=1$. Note that all analysis in this
paper assumes zero rod thickness; in Figure~\ref{fig:typical} the rod
has been artificially fattened for better viewing.

%\bibliography{refs_no_notes}
%\bibliographystyle{plain}

\clearpage
\address{G.~H.~M.~van der Heijden,\\
Centre for Nonlinear Dynamics,\\
University College London,\\
Gower Street,\\
London WC1E 6BT, UK,\\
\email{ucesgvd@ucl.ac.uk}\\
\and
M.~A.~Peletier,\\
Dept.~of Mathematics and Computing Science,\\
Technical University Eindhoven,\\
PO Box 513,\\
5600 MB Eindhoven,\\
The Netherlands,\\
\email{mpeletie@win.tue.nl}\\
\and
R.~Planqu\'e,\\
Centrum voor Wiskunde en Informatica,\\
Kruislaan 413,\\
1098 SJ Amsterdam,\\
The Netherlands,\\
\email{rplanque@cwi.nl}\\
(Corresponding author)}

\end{document}